\documentclass[12pt,a4paper]{article}

\usepackage{latexsym,amssymb}
\usepackage{graphicx,color}
\usepackage{amsmath}
\newcommand\sect[1]{\section{#1}\setcounter{equation}0}

\newcommand\no{\nonumber\\{}}

\newcommand\eqnb{\begin{eqnarray}}
\newcommand\eqne{\end{eqnarray}}

\newcommand{\mf}{\mathfrak}

\newcommand{\C}{\ensuremath{\mathbb{C}}}
\newcommand{\R}{\ensuremath{\mathbb{R}}}
\newcommand{\Z}{\ensuremath{\mathbb{Z}}}

\usepackage{latexsym}

\newcommand\void[1]{}   

\newcommand{\al}{\alpha}

\newcommand{\hs}{\hspace*}

\newcommand{\Tr}{{\mathrm{Tr}}}
\newcommand{\ii}{{\mathrm{i}}}
\newcommand{\sign}{{\mathrm{sign}}}

\newtheorem{theorem}{Theorem}

\newtheorem{lemma}{Lemma}
\newtheorem{proposition}{Proposition}

\begin{document}

{\center{{\huge Gauged non-compact WZNW models, revisited}\vspace{15mm}\\
Jonas Bj\" ornsson\footnote{j.bjornsson@damtp.cam.ac.uk} and Stephen 
Hwang\footnote{stephen.hwang@kau.se} \\
\vspace{.5cm}
$\mathrm{{}^1}$Department of Applied Mathematics and Theoretical Physics\\
Cambridge University\\
Wilberforce Road\\
Cambridge CB3 0WA, United Kingdom\\
\vspace{.5cm}
$\mathrm{{}^2}$Department of Physics\\
Karlstad University \\
Universitetsgatan 2\\
SE-651 88 Karlstad, Sweden}\vspace{15mm} \\}


\begin{abstract}
The purpose of the present paper is to investigate the necessary conditions for unitarity of the spectrum of non-compact gauged WZNW models to some depth. In particular, we would like to investigate the necessity of integer weights and level. We will learn that the problem is very complex and we have not found any simple and general way to formulate the necessary conditions. Instead one must resort to studying the problem almost case by case. The only nearly complete conditions that we will find, is for the case $\mf{g} = \mf{su}(n,1)$. Furthermore, the horizontal part of the case $\mf{g} = \mf{su}(p,q)$ is nearly completed as well. In other cases, we will find conditions associated with certain subalgebras and nodes in the Dynkin diagram close to the one corresponding to the non-compact root. In these examples we can give conditions for the horizontal part of the algebra. As a by-product of our investigation we will prove some nice formulae of character identities and explicit branching functions for these representations.\end{abstract}


\sect{Introduction}

String theory propagating on space-time manifolds with time not being part of a Minkowski space is a subject of importance.  In the beginning, \cite{Balog:1988jb}--\cite{Henningson:1991jc}, the interest was motivated by the fact that more general non-compact backgrounds than Minkowski space was most likely to be inevitable, if one considered a theory that included gravity. More recently the insight into the non-perturbative properties of string theory leading to M-theory, the non-compact backgrounds associated with Anti de Sitter (AdS) spaces became of central interest \cite{Maldacena:1997re}. Although the progress has been substantial, knowledge of the properties beyond low-energy or semi-classical solutions to string theory for these backgrounds is limited.

In \cite{Hwang:1998tr}--\cite{Bjornsson:2008fq} string theories connected to non-compact Hermitian symmetric spaces were considered. These theories are constructed as gauged $G/H'$ WZNW models, where $H=H'\times Z\left(H\right)$, with $Z\left(H\right)$ being the one-dimensional center of $H$, is the maximal compact subgroup of $G$. The spaces $G/H$ are associated with non-compact Hermitian symmetric spaces. All such possible spaces have been classified (see \cite{Helgason} and references therein). There are four large classes\footnote{Due to isomorphisms between different Lie algebras, we only need to treat the cases $p\geq5$ for $SO(p,2)$, $n\geq5$ for $SO^\ast (2n)$}: $G= SU(p,q)$ with $p,q\geq1$, $SO(p,2)$ with $p\geq3$, $SO^\ast (2n)$ with $n\geq2$ and $Sp(2n,\mathbb{R})$ with $n\geq1$, and, in addition, two cases associated with real forms of the exceptional groups $E_6$ ($E_{6|-14}$) and $E_7$ ($E_{7|-25}$). The corresponding string theories studied in \cite{Hwang:1998tr} are then represented as WZNW models associated with the cosets $G/H'$. Time-like field components in this model are represented by current modes associated with $Z\left(H\right)$. $Z\left(H\right)$ is topologically a circle, therefore, to get more realistic string models one should consider the infinite covering  $S^1\rightarrow \mathbb{R}$. For the purpose considered here, this is not particularly important, as the question of the unitarity of the state-space does not depend on this. 

In \cite{Bjornsson:2007ha} we formulated in Theorem 1 necessary and sufficient conditions for unitarity for classes of highest weight representations. The results stated there, when it comes to necessary conditions, are not correct for $G$ of rank three and higher,  as was discussed in \cite{Bjornsson:2008fq}. The purpose of the present paper is to investigate the necessary conditions to some depth. In particular, we would like to investigate the necessity of integer weights and level. We will learn that the problem is very complex and we have not found any simple and general way to formulate the necessary conditions. Instead one must resort to studying the problem almost case by case. The only nearly complete conditions that we will find, is for the case $\mf{g} = \mf{su}(n,1)$. Furthermore, the horizontal part of the case $\mf{g} = \mf{su}(p,q)$ is nearly completed as well. In other cases, we will find conditions associated with certain subalgebras and nodes in the Dynkin diagram close to the one corresponding to the non-compact root. In these examples we can give conditions for the horizontal part of the algebra. As a by-product of our investigation we will prove some nice formulae of character identities and explicit branching functions for these representations. 

The paper is organized as follows. The second section is devoted to the formalism of the paper and a short review of paper \cite{Bjornsson:2007ha}. In the section following we discuss the unitarity in general for these models and a decomposition of the character w.r.t.\ a subalgebra is proven for certain weights. In the third section the algebra $A_{r_{\mf{g}}}$ is treated and in the fourth section conditions from embedded submodules are discussed. In the fifth section, the necessary conditions associated to embedded submodules are investigated. The last section is devoted to a short discussion. In appendices A and B we prove Proposition \ref{characterexpansion} and Theorem \ref{charsigntheorem}, respectively.


\sect{The formalism}

This paper relies heavily on the formalism and notation introduced in \cite{Bjornsson:2007ha}\footnote{Observe that equation, lemma and theorem numbers refers to the hep-th version of the paper.} (which we henceforth will refer to as the "previous paper"). It is based on the construction of the gauged WZNW model through the BRST formalism developed in \cite{Karabali:1989dk}. All notation and reference to lemmas etc.\ are found in the previous paper. We will here only consider the bosonic coset model, but generalizing our results to the world-sheet supersymmetric model \cite{Schnitzer:1988qj}--\cite{FigueroaO'Farrill:1995pv}, considered in \cite{Bjornsson:2008fq} is straightforward. This follows since the condition that the highest weight is antidominant implies that one only has to study the submodules corresponding to the bosonic creation operators, as the fermions give a unitary spectrum. We will explain in more detail below why the proof of necessary conditions formulated in Theorem 1 in the previous paper fails. But it should be noted that all other results in the above paper .eg. Theorem 2 and Corollary 1 hold as formulated and proven. 

The state-space of the gauged $G/H$ WZNW model\footnote{Whenever we in the paper refer to the state-space without further specifications, then we mean this state-space.} is defined through the BRST condition. Denoting by $Q$ the BRST charge, the physical state-space of the string theory is defined by the conditions
\eqnb
Q\left| \Phi\right>
      &=&
          0
\no
b^i_0\left| \Phi\right>
      &=&
          0\phantom{1234}i=2,\ldots,r_{\mf{g}}
\no
\mathcal{P}_0\left| \Phi\right>
      &=&
          0.
\label{brstcondition}
\eqne
We denote by
${\cal H}^{Q}_{\hat{\mu}\tilde{\hat{\mu}}}$ the subspace of states of $\mathcal{H}_{\hat\mu}^{\hat{\mf{g}}} \times\tilde{\mathcal{H}}^{\hat{\mf{h}}'}_{\hat{\tilde{\mu}}}\times\mathcal{H}^{\mathrm{CFT}}_{l'}\times\mathcal{H}^{ghost}$ satisfying the above equations and being $Q$ non-exact. Here $\mathcal{H}^{\mathrm{CFT}}_{l'}$ represents some unitary CFT. $ b^i_0$ and $\mathcal{P}_0$ are ghost-momenta zero modes.

The representations that we focused on in the previous paper, and will focus on here,  are antidominant highest weight representations
for $\hat{\mf{g}}^\C$.  We also require that there should exist a BRST invariant ground-state i.e. a state of the form
\eqnb
\left| 0;\mu,\tilde \mu\right> 
      &\equiv&
          \left| 0;\mu\right> \otimes \left| \tilde 0;\tilde \mu\right> \otimes \left| 0\right>_{ghost}.
\eqne
This state is a non-trivial BRST invariant state if $\mu^i+\tilde {\mu}^i+2\rho_{\mf{h}'}^i=0$, $i=2,\ldots, r_{\mf{g}}$. Therefore, if we choose $\mu$ to be antidominant we must require $\tilde{\mu}$ to be a dominant ${{\mf{h}}'}^\C$ weight.  We remind also of Lemma 1 of the previous paper which states that for non-trivial states we must have $(\mu^i+\tilde {\mu}^i)\in\mathbb{Z}$, $i=2,\ldots, r_{\mf{g}}$. This implies, in particular, that if $\tilde {\mu}^i$ is an integer so is $\mu^i$. Furthermore, $\tilde \kappa = -\kappa-2g^{\vee}_{\mf{h}'}$ so that if $\tilde \kappa$ is an integer then so is also $\kappa$. Here $\kappa=I_{\mf{h}'\subset\mf{g}}k$ where $I_{\mf{h}'\subset\mf{g}}$ is the Dynkin index of the embedding defined as
\eqnb
I_{\mf{h}'\subset\mf{g}} &=& \frac{\left(\theta\left(\mf{g^\C}\right),\theta\left(\mf{g^\C}\right)\right)}{\left(\theta\left(\mf{h'^\C}\right),\theta\left(\mf{h'^\C}\right)\right)},
\eqne
where $\theta\left(\cdot\right)$ is the highest root in each algebra expressed in the root system of $\mf{g}^{\C}$.

The necessary conditions on the state-space that are the focus of this paper are formulated as conditions on the dominant highest weights $\hat{\tilde{\mu}}$ such that state-space will be unitary for {\it any} antidominant highest weight $\hat{\mu}$.  Note that the necessary conditions were formulated in the previous paper in a slightly different way. We believe that the present formulation is more natural from the point of view of interactions and modular invariance.

One of the central points in the analysis of unitarity in the previous paper, and will be in the investigation here, is Lemma 2 in the previous paper. This lemma states that we get necessary and sufficient conditions for unitarity from solutions of the equations
\eqnb
\int d\tau \left[\mathcal{B}^{\left(\hat{\mf{g}},\,\hat{\mf{h}}'\oplus Vir\right)}(\tau,\phi)-
\mathcal{S}^{\left(\hat{\mf{g}},\,\hat{\mf{h}}'\oplus Vir\right)}(\tau,\phi)\right]=0.
\label{eq23}
\eqne
where $\mathcal{B}$ and $\mathcal{S}$ are the generalized branching function and coset signature function, respectively. By inserting the explicit expressions for them for the known sectors, namely the $\hat{\mf{g}}$- and ghost sectors one arrives at the equation (4.32) in the previous paper:
\eqnb
0
      &=&
          \int d\tau d\theta\,\left[
          \chi^{\left(\hat{\mf{g}},\,\hat{\mf{h}}'\oplus Vir\right)}(\tau,\phi,\theta)-
          \,\Sigma^{\left(\hat{\mf{g}},\,\hat{\mf{h}}'\oplus Vir\right)}
          (\tau,\phi,\theta)\right]
      \no
      &=&
          \int d\tau d\theta\, e^{-2\pi\ii \tau}\chi^1_\mu(\tau,\phi,\theta)
          \left[\chi_{\tilde\mu}^{\hat{\mf{h}}'}(\tau, \theta)-
          \Sigma^{\hat{\mf{h}}'}_{\tilde\mu}(\tau, \theta)\right],
\eqne
where $d\theta \equiv \prod_{l=2}^{r_\mf{g}}d\theta^{l}$ are formal integrations projecting onto the $\theta$-independent part of the integrand. We will throughout this paper assume that the mass-shell condition may be solved without imposing further conditions. This can be motivated in part by requiring that unitarity should hold if we add an arbitrary unitary conformal field theory. Furthermore, if we were to take into account the mass-shell condition, it would complicate our analysis considerably. Assuming a solution of the mass-shell condition implies that the $\tau$-integration above is trivial and can be dropped. Then the above equation reads
\eqnb
0     &=&
          \int d\theta\, e^{-2\pi\ii \tau}\chi^1_\mu(\tau,\phi,\theta)
          \left[\chi_{\tilde\mu}^{\hat{\mf{h}}'}(\tau, \theta)-
          \Sigma^{\hat{\mf{h}}'}_{\tilde\mu}(\tau, \theta)\right].
\label{chi-sigma}
\eqne
Here $\chi^1_\mu(\tau,\phi,\theta)$ is given by eq.~(4.31) in the previous paper (where we drop the notation $\alpha_\parallel$
since $(\theta,\alpha_\parallel)=(\theta,\alpha)$),
\eqnb
\chi^1_\mu(\tau,\phi,\theta) 
      &\equiv&
          q^{\frac{\mathcal{C}_2^{\mf{g}}(\mu)}{2\left(k+g_{\mf{g}}^\vee\right)}} 
          e^{\ii\phi\mu_1+\ii\left(\theta,\mu+2\rho_{\mf{h}'}\right)}
      \no
      &\times&
          \prod_{\alpha\in\Delta^+_c}\left(1-e^{-\ii\left(\theta,\alpha\right)}\right)
          \prod_{\alpha\in\Delta^+_n}\frac{1}{1-e^{-\ii\phi-\ii\left(\theta,\alpha\right)}}
      \no
      &\times&
          \prod_{m=1}^{\infty}
          \left\{
             \left(1-q^m\right)^{r_{\mf{g}}}
             \prod_{\alpha\in\Delta_c}\left(1-q^m e^{\ii\left(\theta,\alpha\right)}\right)
          \right.   
          \no
      &\times&
      \left.
          \prod_{\alpha\in\Delta^+_n}\frac{1}{1-q^m e^{-\ii\phi-\ii\left(\theta,\alpha\right)}}
          \prod_{\alpha\in\Delta^+_n}\frac{1}{1-q^m e^{\ii\phi+\ii\left(\theta,\alpha\right)}}
          \right\},
\label{char-tot}
\eqne
where $q \equiv \exp\left[2\pi\ii\tau\right]$.

Before we proceed further in our investigation, let us briefly discuss why the proof of necessary conditions in the previous paper failed. The logic of the proof was to first establish results for all rank two cases, Lemma 4, and then to use this result for the general case by looking at rank two subalgebras of $\mf{g}^\C$. The connection between the rank  two case and the general case was made through a gradation, eq.~(4.52), and a decomposition of the BRST operator, eq.~(4.53). It was established through the gradation that any non-trivial BRST invariant state is also invariant w.r.t.\ the BRST operator connected to the rank two subalgebra. However, which was wrongly assumed, this does not imply that necessary conditions for unitarity of the state-space of the rank two subspace give necessary conditions for the full space.


\sect{Unitarity}

We now proceed with our investigation. We introduce the expansion, eq.~(4.33) of the previous paper,
\eqnb
\chi^{{\hat{\mf{h}}}'}_{\tilde\mu}\left(q,\theta\right)-
\Sigma_{\tilde\mu}^{{\hat{\mf{h}}}'}\left(q,\theta\right)
      &=&
          q^{-\frac{\mathcal{C}_2^{\tilde{\mf{h}}'}(\tilde\mu)}{2\left(\kappa+g_{\mf{h}'}^\vee\right)}}
          e^{\ii(\theta,\tilde\mu)}\sum_{n=0}^\infty \sum_{\tilde\lambda\in{\Gamma_r^{\mathfrak{h}'}}^+}N_{n,\tilde\lambda}
          q^{n}e^{-\ii(\theta, \tilde\lambda-n\theta'_{\mf{h}'})},
\label{expansion1}
\eqne
where $N_{n,\tilde\lambda}$ is twice the number of negatively normed states at grade $n$ and weight $\tilde{\mu}-\tilde{\lambda}$. $\theta'_{\mf{h}'}$ is the highest root of $\mf{h}^{\prime\mathbb{C}}$. In studying eq.~(\ref{chi-sigma}), using eq.~(\ref{char-tot}), a key observation is that, apart from the explicit exponential factor, the character in eq.~(\ref{chi-sigma}) is independent of $\hat{\mu}$, as long as it is antidominant. On the other hand, this is not true for the auxiliary sector, where the unknown coefficients $N_{n,\tilde\lambda}$ depend crucially on $\hat{\tilde{\mu}}$. This will simplify our analysis as we take $\hat{\tilde{\mu}}$ to be fix and consider different values of $\hat{\mu}$ during our analysis.

We will in the following mainly concentrate on the grade zero part of the state-space i.e. the $q$-independent part of eq.~(\ref{chi-sigma})
\eqnb
\int d\theta e^{\ii\left(\theta,\tilde\mu-n_i\alpha^{(i)}\right)}\chi_{\mu}^1(\phi,\theta)N_{n_i\alpha^{(i)}}
	&=&
		0
\label{Integration:horizontal}
\eqne
where $\chi_{\mu}^1(\phi,\theta)$ is the horizontal (i.e. grade zero) part of eq.\ (\ref{char-tot}),
\eqnb
\chi_{\mu}^1(\phi,\theta) 
	&=&
		e^{\ii\left(\mu + 2\rho_{\mf{\tilde{h}}'},\theta\right)} \frac{\prod_{\alpha\in\Delta_c^+}\left(1-e^{-\ii\left(\theta,\alpha\right)}\right)}{\prod_{\alpha\in\Delta_n^+}\left(1-e^{-\ii\left(\theta,\alpha\right) - \ii\phi}\right)}.
\label{chi1:hori}
\eqne
Due to the factors in the numerator the analysis of the integral is difficult. But we will prove a theorem which simplifies the analysis. 

\begin{theorem}
The character of the irreducible representation, $\mathcal{L}^\sigma_{\mf{g}}$, of $\mf{g}^\C$ of highest weight $\sigma^i$: $\sigma^i=0$, $i=2,\ldots,r_{\mf{g}}$ and $\left(\sigma+\rho_{\mf{g}},\theta_{\mf{g}}\right)<0$, can be written as
\eqnb
\chi_\sigma^{\mf g}(\phi,\theta)&=& \frac{e^{\ii\sigma_1\phi}}{  \prod_{\alpha\in\Delta^+_n}
\left(1-e^{-\ii\phi-\ii\left(\theta,\alpha\right)}\right)},
\label{charg.simple}
\eqne
or as
\eqnb
\chi_\sigma^{\mf g}(\phi,\theta)&=& e^{\ii\sigma_1\phi}\sum_{m=0}^\infty\ e^{-\ii m\phi}
\sum_{\lambda\in C_m\cap\Gamma^+_{\mf{h}'}} c_{\lambda,m}\hs{1mm}\chi_\lambda^{\mf{h}'}(\theta),
\label{branchingfuncdecomp}
\eqne
where
\eqnb
\chi_\lambda^{\mf{h}'}(\theta) &\equiv&
\frac{\sum_{w\in W_{\mf{h}'}}\mathrm{sign}(w)e^{\ii(\theta,w(\lambda_m+\rho_{\mf{h}'})-\rho_{\mf{h}'})}}
{ \prod_{\alpha\in\Delta^+_c}\left(1-e^{-\ii\left(\theta,\alpha\right)}\right)}. 
\eqne
Here $C_m=\{\lambda:\lambda=I_{\mf{h}'\subset\mf{g}}m\Lambda_{(2)}-n_i\alpha^{(i)}\;n_i\in\Z_+\}$. $c_{\lambda,m}\in \mathbb{Z}_+$ and, in particular, $c_{\lambda_m,m}=1$ where $\lambda_m\equiv -m\al^{(1)}_\parallel$.
\label{theorem1}
\end{theorem}

Note that eq.\ (\ref{branchingfuncdecomp}) is a necessary condition for $\cal{L}_{\mf{g}}^\sigma$ to be completely reducible w.r.t.\ $\mf{h}'^\C$ with highest weights in $C_m\cap\Gamma^+_{\mf{h}'}$. The constants $c_{\lambda,m}$ are, as will be shown below, the number of highest weight states w.r.t. $\mf{h}'^\C$ of weight $\lambda$ for a given $m$. $W_{\mf{h}'}$ is the Weyl group of $\mf{h}'^\C$.

\paragraph{Proof:}
The first result, eq.~(\ref{charg.simple}), follows from the fact that any state, for the given highest weight $\sigma$ and of the form
\eqnb
\prod_{i=2}^{r}\left(E^{-\alpha^{(i)}}\right)^{n_i}\left|\sigma\right>,
\label{eq212}
\eqne
is a null-state. It is straightforward to establish that the remaining states give the character  eq.~(\ref{charg.simple}) (this is a special case of the characters presented in chapter 7.4 in \cite{jonas}). 

We now proceed to prove the remaining claims of the theorem. We expand the denominator as
\eqnb
\frac{1}{\prod_{\alpha\in\Delta_n^+}\left(1-e^{-\ii\left(\theta,\alpha\right)-\ii\phi}\right)}
	&=&
		\sum_{m=0}^{\infty}\sum_{\mu\in\Gamma_{\mf{h}'}}c_{\mu,m}e^{-\ii m\phi}\frac{e^{\ii \left(\mu+\rho_{\mf{h}'},\theta\right)}}{e^{\ii(\theta,\rho_{\mf{h}'})}\prod_{\alpha\in\Delta_c^+}(1-e^{-\ii\left(\theta,\alpha\right)})},
\no
\eqne
where $c_{\mu,m}$ are integer coefficients and $\Gamma_{\mf{h}'}$ is the fundamental weight lattice of $\mf{h}'^\C$. The left-hand side of the above equation is Weyl-invariant w.r.t.\ the subalgebra. The factor $e^{\ii(\theta,\rho_{\mf{h}'})}\prod_{\alpha\in\Delta_c^+}(1-e^{-\ii\left(\theta,\alpha\right)})^{-1}$ is skew-symmetric under such Weyl transformations. Applying a Weyl transformation yields
\eqnb
\sum_{m=0}^{\infty}\sum_{\mu\in\Gamma_{\mf{h}'}}c_{\mu,m}e^{-\ii m\phi}\frac{e^{\ii \left(\mu+\rho_{\mf{h}'},\theta\right)}}{e^{\ii(\theta,\rho_{\mf{h}'})}\prod_{\alpha\in\Delta_c^+}(1-e^{-\ii\left(\theta,\alpha\right)})}
&=&
\no
\sign(w)\sum_{m=0}^{\infty}\sum_{\mu\in\Gamma_{\mf{h}'}}c_{\mu,m}e^{-\ii m\phi}\frac{e^{\ii \left(w(\mu+\rho_{\mf{h}'}),\theta\right)}}{e^{\ii(\theta,\rho_{\mf{h}'})}\prod_{\alpha\in\Delta_c^+}(1-e^{-\ii\left(\theta,\alpha\right)})}.
\eqne
The invariance under Weyl transformations implies
\eqnb
c_{\mu,m} &=& \sign(w)c_{w(\mu+\rho_{\mf{h}'})-\rho_{\mf{h}'},m}.
\eqne
As one always can find a Weyl transformation that takes a weight to the fundamental Weyl chamber, one only needs to sum over all roots in the positive root lattice and the Weyl group of the subalgebra. Thus, the character can be expanded as
\eqnb
\frac{1}{\prod_{\alpha\in\Delta_n^+}\left(1-e^{- \ii\left(\theta,\alpha\right) - \ii\phi}\right)} 
	&=&
		\sum_{m=0}^{\infty}\sum_{\mu\in\Gamma^+_{\mf{h}'}}c_{\mu,m}e^{-\ii m\phi}\frac{\sum_{w\in W_{\mf{h}'}}\sign(w)e^{\ii \left(\theta,w(\mu+\rho_{\mf{h}'})-\rho_{\mf{h}'}\right)}}{\prod_{\alpha\in\Delta_c^+}\left(1-e^{-\ii\left(\theta,\alpha\right)}\right)},
\label{Expansion.chi1}
\no
\eqne
where ${c}_{\mu,m}$ are integers depending on $m$ and $\mu$. In order to prove that  ${c}_{\mu,m}$ are positive integers, we use the BRST formalism and apply it to $\cal{L}_{\mf g}^\sigma$ gauging the subalgebra $\mf{h}'^\C$. The BRST charge is the grade zero part of eq.~(3.2) in the previous paper. As the weights of $\mf{h}'^\C$ arising in the sum in eq.\ (\ref{Expansion.chi1}) are dominant, the representations for the auxiliary sector satisfy $\tilde\sigma \leq -2\rho_{\mf{h}'}$. Thus, they are antidominant giving the corresponding character to be the one of the Verma module. The ghost sector has a character which is the grade zero part of the previous paper, eq.~(4.27). Using these characters together with the one in eq.\ (\ref{Expansion.chi1}) and integrating over $\theta$ one finds (cf.\ the proof of Proposition 1 in \cite{Hwang:1994yr}) 
\eqnb
\int d\theta\ \chi_\sigma^{\mf g}(\phi,\theta)\chi_{\tilde\sigma}^{\mf{h}'}(\theta)\chi^{\mathrm{gh}}
&=& e^{\ii\sigma_1\phi}\sum_{m=0}^\infty\ 
\sum_{\lambda\in \Gamma_{\mf{h}'}^+ }c_{\lambda,m}e^{-\ii m\phi}.
\label{eq215}
\eqne
This shows that the constants $c_{\lambda,m}$ are the number of non-trivial BRST invariant states at the weight  $\lambda$ for a given $m$. Consequently, $c_{\lambda,m}\in\mathbb{Z}_+$. In addition, by applying the results of \cite{Hwang:1993nc}, which are valid here as the auxiliary sector has antidominant highest weights since $\mu$ is dominant, we know that the non-trivial BRST invariant states satisfy
\eqnb
E^{\alpha^{(i)}}\left|\mu\right>=0, \hs{3mm}i=2,\ldots , r_{\mf{g}},
\label{eq216}
\eqne
i.e.\ are highest weight states w.r.t. $\mf{h}'^\C$. 

We can now consider restrictions of the sum over the positive weight lattice. We can write ${c}_{\mu,m} = \delta_{I_{\mf{h}'\subset\mf{g}}m\Lambda_{(2)},\mu} + {c}'_{\mu,m}$, where ${c}'_{\mu,m} \neq 0$ only when $\mu$ is of the form $\mu = I_{\mf{h}'\subset\mf{g}}m\Lambda_{(2)}-n_i\alpha^{(i)}$ where $n_i\geq 0$ but not all zero. This is due to the fact that $\left(E^{-\alpha^{(1)}}\right)^m\left|\mu\right>$ is a highest weight state w.r.t.\ the subalgebra $h'^\C$ and, for any fixed $m$, the highest possible weight. Thus, one gets
\eqnb
\frac{1}{\prod_{\alpha\in\Delta_n^+}\left(1-e^{- \ii\left(\theta,\alpha\right)- \ii\phi}\right)} 
	&=&
		\sum_{m=0}^{\infty} \sum_{\mu\in\Gamma^+_{\mf{h}'} \cap C_m}c_{\mu,m}e^{-\ii m\phi}\frac{\sum_{w\in W_{\mf{h}'}}\sign(w)e^{\ii \left(\theta,w(\mu+\rho_{\mf{h}'})-\rho_{\mf{h}'}\right)}}{\prod_{\alpha\in\Delta_c^+}\left(1-e^{-\ii\left(\theta,\alpha\right)}\right)}
\label{Decompositionh}
\no
\eqne
where $C_m=\{\lambda:\lambda = I_{\mf{h}'\subset\mf{g}}m\Lambda_{(2)}-n_i\alpha^{(i)}\;n_i\in\Z_+\}$. We have now proven the theorem. $\Box$

In a few cases one can determine $c_{\mu,m}$. These are summarized in the following proposition.
\begin{proposition}
We have in the following three cases
\paragraph{1. $\mf{g}^\C = A_{p+1}$ and $\mf{h}'^\C = A_p$:} 
$c_{\mu,m} = \delta_{\mu,m\Lambda_{(2)}}$ and
\eqnb
\frac{1}{\prod_{\alpha\in\Delta_n^+}\left(1-e^{- \ii\left(\theta,\alpha\right)- \ii\phi}\right)} 
	&=&
		\sum_{m=0}^{\infty}e^{- \ii m\phi}\frac{\sum_{w\in W_{\mf{h}'}}\sign(w)e^{\ii\left(\theta,w(m\Lambda_{(2)}+\rho_{\mf{h}'})-\rho_{\mf{h}'}\right)}}{\prod_{\alpha\in\Delta_c^+}\left(1-e^{-\ii\left(\theta,\alpha\right)}\right)}
\eqne
\paragraph{2.~ $\mf{g}^\C = B_{p+1}$ and $\mf{h}'^\C = B_p$:}

$c_{\mu,m} = \sum_{n=0}^{\left[\frac{m}{2}\right]}\delta_{\mu,(m-2n)\Lambda_{(2)}}$ and
\eqnb
\frac{1}{\prod_{\alpha\in\Delta_n^+}\left(1-e^{- \ii\left(\theta,\alpha\right)- \ii\phi}\right)} 
	&=&
		\frac{1}{1-e^{-2\ii\phi}}\sum_{m=0}^{\infty}e^{-\ii m\phi}\frac{\sum_{w\in W_{\mf{h}'}}\sign(w)e^{\ii \left(\theta,w(m\Lambda_{(2)}+\rho_{\mf{h}'})-\rho_{\mf{h}'}\right)}}{\prod_{\alpha\in\Delta_c^+}\left(1-e^{-\ii\left(\theta,\alpha\right)}\right)}
\no
\eqne
\paragraph{3.~ $\mf{g}^\C = D_{p+1}$ and $\mf{h}'^\C = D_p$:} 
$c_{\mu,m} = \sum_{n=0}^{\left[\frac{m}{2}\right]}\delta_{\mu,(m-2n)\Lambda_{(2)}}$ and 
\eqnb
\frac{1}{\prod_{\alpha\in\Delta_n^+}\left(1-e^{- \ii\left(\theta,\alpha\right)- \ii\phi}\right)} 
	&=&
		\frac{1}{1-e^{-2\ii\phi}}\sum_{m=0}^{\infty}e^{-\ii m\phi}\frac{\sum_{w\in W_{\mf{h}'}}\sign(w)e^{\ii\left(\theta,w(m\Lambda_{(2)}+\rho_{\mf{h}'})-\rho_{\mf{h}'}\right)}}{\prod_{\alpha\in\Delta_c^+}\left(1-e^{-\ii\left(\theta,\alpha\right)}\right)}
\no
\eqne
\label{characterexpansion}
\end{proposition}
The proof of this proposition is given in Appendix A. 


\sect{Necessary conditions for unitarity for $\mf{g}^\C = A_{r_{\mf{g}}}$}

In this section we will discuss the simplest example of general rank, the example of $\mf{g}^\C = A_{r_{\mf{g}}}$. For this case we will be able to establish quite general necessary conditions for unitarity. In doing this we will introduce some general methods which we believe will generalize to other algebras, as well. We will here focus on representations we call strictly dominant representations for the auxiliary sector. These representations are defined as
\eqnb
\tilde{\mu}^i &>& -1 \phantom{1234} \mathrm{for}\; i=2,\ldots,r_\mf{g}\no
\tilde{k}     &>& \left(\tilde{\mu}^i,\theta'_{\mf{h}}\right) \no
\tilde{k}     &>& -g^{\vee}_{\mf{h}'}. 
\eqne

First we will establish the lemma.
\begin{lemma}
Let $\hat{\tilde{\mu}}$ be a strictly dominant highest weight and $3\leq i\leq r_{\mf{g}}$. If $\tilde{\mu}^{j}\in \Z_+$ for $2\leq j\leq i-1$, $\tilde{\mu}^{i-1}\geq 1$ and $\tilde{\mu}^{i} \notin \Z_+$, then there exists at least one antidominant highest weight $\hat{\mu}$ such that the state-space is non-unitary.
\label{lemma1}
\end{lemma}
In proving this lemma we also will prove 
\begin{lemma}
Let $\hat{\tilde{\mu}}$ be a strictly dominant highest weight.  If $\tilde{\mu}^{2} \notin \Z_+$ then there exists at least one antidominant highest weight $\hat{\mu}$ such that the state-space is non-unitary.
\label{lemma2}
\end{lemma}
{\bf Proof:} To prove these two lemmas we use a theorem which we prove in Appendix B.

\begin{theorem}
Let $\tilde{\mu}$ be a strictly dominant highest weight satisfying $\tilde{\mu}^j\in\Z_+$ for $j=2,\ldots,i-1$, $3\leq i\leq r_{\mf{g}}$, then
\eqnb
\chi^{\mf{h}'}_{\tilde{\mu}}\left(\theta\right) 
    &=&
        e^{\ii\left(\tilde{\mu},\theta\right)}\frac{\sum_{w\in W_{i-2}}\sign(w)e^{\ii\left(w(\tilde{\mu}+\rho_{\mf{h}'})-\tilde{\mu}-\rho_{\mf{h}'},\theta\right)}}{\prod_{\alpha\in\Delta^{+}_c}\left(1-e^{-\ii\left(\alpha,\theta\right)}\right)}
    \no
    &-&
        \delta_{\tilde{\mu}^i,\left[\tilde{\mu}^i\right]}e^{\ii\left(\tilde{\mu},\theta\right)}\frac{\sum_{w\in W_{i-2}}\sign(w)e^{\ii\left(w\left(\tilde{\mu}+\rho_{\mf{h}'}-\left(\left[\tilde{\mu}^{i}\right]+1\right)\alpha^{(i)}\right)-\tilde{\mu}-\rho_{\mf{h}'},\theta\right)}}{\prod_{\alpha\in\Delta^{+}_c}\left(1-e^{-\ii\left(\alpha,\theta\right)}\right)}
    \no
    &+&
        \mathcal{O}\left(e^{\ii\left(\gamma,\theta\right)}\right)
\\
\Sigma^{\mf{h}'}_{\tilde{\mu}}\left(\theta\right) 
    &=&
        e^{\ii\left(\tilde{\mu},\theta\right)}\frac{\sum_{w\in W_{i-2}}\sign(w)e^{\ii\left(w(\tilde{\mu}+\rho_{\mf{h}'})-\tilde{\mu}-\rho_{\mf{h}'},\theta\right)}}{\prod_{\alpha\in\Delta^{+}_c}\left(1-e^{-\ii\left(\alpha,\theta\right)}\right)}
        \no
    &-&
        \delta_{\tilde{\mu}^{i},\left[\tilde{\mu}^{i}\right]}e^{\ii\left(\tilde{\mu},\theta\right)}\frac{\sum_{w\in W_{i-2}}\sign(w)e^{\ii\left(w\left(\tilde{\mu}+\rho_{\mf{h}'}-\left(\left[\tilde{\mu}^{i}\right]+1\right)\alpha^{(i)}\right)-\tilde{\mu}-\rho_{\mf{h}'},\theta\right)}}{\prod_{\alpha\in\Delta^{+}_c}\left(1-e^{-\ii\left(\alpha,\theta\right)}\right)}
    \no
    &-&
        2\left(1-\delta_{\tilde{\mu}^{i},\left[\tilde{\mu}^{i}\right]}\right)e^{\ii\left(\tilde{\mu},\theta\right)}\frac{\sum_{w\in W_{i-2}}\sign\left(w\right)e^{\ii\left(w\left(\tilde{\mu}+\rho_{\mf{h}'}-\left(\left[\tilde{\mu}^{i}\right]+2\right)\alpha^{(i)}\right)-\tilde{\mu}-\rho_{\mf{h}'},\theta\right)}}{\prod_{\alpha\in \Delta_+^{c}}\left(1-e^{-\ii\left(\alpha,\theta\right)}\right)}
    \no
    &+&
        \mathcal{O}\left(e^{\ii\left(\gamma,\theta\right)}\right),
\eqne
where $W_{i-2}$ denotes the subgroup of $W_{\mf{h}'}$ generated by $w_{(j)}$ for $j=2,\ldots,i-1$ and $\gamma$ satisfies
\eqnb
\left(\mu-\gamma,\Lambda_{(j)}\right)
&>&
\left\{
\begin{array}{ll}
\tilde{\mu}^i+1              & j=i\phantom{12} \tilde{\mu}^i\in\Z_+\\
\left[\tilde{\mu}^i\right]+2 & j=i\phantom{12} \tilde{\mu}^i\notin\Z_+ \\
0                            & i+1 \leq j \leq r_\mf{g}
\end{array}
\right.
\eqne
\label{charsigntheorem}
\end{theorem}

From Theorem \ref{charsigntheorem}, we get that the character minus the signature function is
\eqnb
\chi^{\mf{h}'}_{\tilde{\mu}}\left(\theta\right)-\Sigma^{\mf{h}'}_{\tilde{\mu}}\left(\theta\right)
&=&
2\left(1-\delta_{\tilde{\mu}^{i},\left[\tilde{\mu}^{i}\right]}\right)e^{\ii\left(\tilde{\mu},\theta\right)}\frac{\sum_{w\in W_{i-2}}\sign\left(w\right)e^{\ii\left(w\left(\tilde{\mu}+\rho_{\mf{h}'}-\left(\left[\tilde{\mu}^{i}\right]+2\right)\alpha^{(i)}\right)-\tilde{\mu}-\rho_{\mf{h}'},\theta\right)}}{\prod_{\alpha\in \Delta_+^{c}}\left(1-e^{-\ii\left(\alpha,\theta\right)}\right)}
\no
&+&
\mathcal{O}\left(e^{\ii\left(\gamma,\theta\right)}\right).
\eqne
This is the expression for $\sum_{m_i\in\Z_+}N_{m_i\alpha^{(i)}}e^{-\ii\left(\alpha^{(i)},\theta\right)}$. Multiplying this expression with $\chi_{\mu}^1(\phi,\theta)$, given in eq.\ (\ref{chi1:hori}), and inserting the result into eq.\ (\ref{Integration:horizontal}) gives
\eqnb
2\sum_{w\in W_{i-2}}\sign\left(w\right)\left(1-\delta_{\tilde{\mu}^i,\left[\tilde{\mu}^i\right]}\right)\int d\theta \frac{e^{\ii\left(\mu + \rho_{\mf{h}'} + w\left(\tilde{\mu}+\rho_{\mf{h}'}-\left(\left[\tilde{\mu}^{i}\right]+2\right)\alpha^{(i)}\right) , \theta\right)}}{\prod_{\alpha\in\Delta^+_{n}}\left(1-e^{-\ii\left(\alpha,\theta\right)-\ii \phi}\right)} &=& 0.
\label{eq4.5}
\eqne
This expression is valid for states with weights $\lambda: \lambda > \gamma$.

For the case $\mf{g} = \mf{su}(r_\mf{g},1)$ there is a simple expression for the denominator of the equation above,
\eqnb
\prod_{\alpha\in\Delta_n^+}\frac{1}{\left(1-e^{-\ii\left(\alpha,\theta\right)-\ii \phi}\right)}
&=&
\sum_{n_1=0}^\infty\sum_{n_2=0}^{n_1}\ldots\sum_{n_{r_{\mf{g}}}=0}^{n_{r_{\mf{g}}-1}}e^{\ii n_1\left( - \phi + \left(\Lambda_{(2)},\theta\right)\right)}e^{-\ii n_2\left(\alpha^{(2)},\theta\right)}\cdot\ldots\cdot e^{-\ii n_{r_{\mf{g}}}\left(\alpha^{(r_{\mf{g}})},\theta\right)}.\no
\label{eq4.6}
\eqne 
This expression is straightforward to prove by expanding the denominator or by using Theorem \ref{theorem1} and Proposition \ref{characterexpansion}.

Inserting eq.\ (\ref{eq4.6}) into eq.\ (\ref{eq4.5}) gives
\eqnb
&&
2\sum_{w\in W_{i-2}}\sum_{n_1=0}^\infty\sum_{n_2=0}^{n_1}\ldots\sum_{n_{r_{\mf{g}}}=0}^{n_{r_{\mf{g}}-1}}\sign\left(w\right)\left(1-\delta_{\tilde{\mu}^i,\left[\tilde{\mu}^i\right]}\right)e^{-\ii n_1\phi}
\no
&&
\int d\theta e^{\ii\left(\mu + \rho_{\mf{h}'} + w\left(\tilde{\mu}+\rho_{\mf{h}'}-\left(\left[\tilde{\mu}^{i}\right]+2\right)\alpha^{(i)}\right) + n_1\Lambda_{(2)} - \sum_{j=2}^{r_{\mf{g}}} n_j\alpha^{(j)} , \theta\right)}=0.
\label{interestingint}
\eqne
As we do not integrate over $\phi$ one needs that this expression is zero for all $n_1$. If the left-hand side of the equation is non-zero we have non-unitarity. The integration over $\theta$ will be non-zero if the equation
\eqnb
\mu + \rho_{\mf{h}'} + w\left(\tilde{\mu}+\rho_{\mf{h}'}-\left(\left[\tilde{\mu}^{i}\right]+2\right)\alpha^{(i)}\right) + n_1\Lambda_{(2)} 
&=&
\sum_{j=2}^{r_{\mf{g}}}n_j\alpha^{(j)},
\label{unitarityeqn}
\eqne
where $n_1$ is arbitrary and $n_1\geq n_2\geq \ldots \geq n_{r_{\mf{g}}}\geq0$, is solvable only for $w=1$ and 
\eqnb
\mu + \tilde{\mu} + 2\rho_{\mf{h}'} + n_1\Lambda_{(2)} + \gamma
&=&
\sum_{j=2}^{r_{\mf{g}}}n_j\alpha^{(j)},
\label{unitarityeqn2}
\eqne 
do not have a solution for the same $n_1$. 

We will make an ansatz of a solution to eq.\ (\ref{unitarityeqn}), which we will check is antidominant, has no solution to eq. (\ref{unitarityeqn2}) and only gets contributions from $w=1$. Therefore, yields a contradiction to eq. (\ref{interestingint}) and yields a non-unitary state-space. The ansatz we make is
\eqnb
\mu &=& -\left(\tilde{\mu} + 2\rho_{\mf{h}'}\right) + \left(\left[\tilde{\mu}^{i}\right]+2\right)\alpha^{(i)} + \left(\left[\tilde{\mu}^{i}\right]+3\right)\sum_{j=2}^{i-1}\alpha^{(j)} - \left(\left[\tilde{\mu}^{i}\right]+3\right)\Lambda_{(2)}.
\label{ansatzhor}
\eqne
First of all, we need to check that this ansatz yields an antidominant weight
\eqnb
\left(\mu,\alpha^{(j)}\right)
&=&
\left\{
\begin{array}{ll}
-\tilde{\mu}^j-2 & j\neq i+1,i,i-1 \\
-\tilde{\mu}^{i+1}-\left[\tilde{\mu}^i\right]-4 & j=i+1 \\
-\tilde{\mu}^{i}+\left[\tilde{\mu}^i\right]-1 & j=i \\
-\tilde{\mu}^{i-1}-1 & j=i-1
\end{array}
\right.,
\eqne
which is antidominant as we have assumed that $\mu^{i-1}\geq 1$. We also have to show that $k < \left(\mu,\theta'_{\mf{g}}\right)$, where $\theta'_{\mf{g}}$ is the highest root of $\mf{g}^\C$. Using $k + \tilde{k} + 2g^{\vee}_{\mf{h}'}$ and the ansatz in eq.\ (\ref{ansatzhor}), the inequality yields
\eqnb
\tilde{k} + 1 + \mu^1 &>& \left(\tilde{\mu},\theta'_\mf{h'}\right),
\eqne
if $i=2$;
\eqnb
\tilde{k} + 2 + \mu^1 &>& \left(\tilde{\mu},\theta'_\mf{h'}\right),
\eqne
if $3\leq i\leq r_\mf{g}-1$. If $i= r_\mf{g}$ the ansatz has to satisfy 
\eqnb
\tilde{k} + 4 + \mu^1 + \left[\tilde{\mu}^{r_{\mf{g}}}\right]&>& \left(\tilde{\mu},\theta'_\mf{h'}\right).
\eqne
As $\mu^1<-1$, we can always choose $\mu$ such that $\hat\mu$ is antidominant in all cases.

Inserting eq.\ (\ref{ansatzhor}) into eq.\ (\ref{unitarityeqn}) for $w=1$ yields that $n_1=\left[\tilde{\mu}^{i}\right]+3$ and the sum will only have contributions from
\eqnb
n_j
&=&
\left\{
\begin{array}{ll}
\left[\tilde{\mu}^{i}\right]+3 & 2 \leq j \leq i-1  \\
0                              & i \leq j \leq r_{\mf{g}}
\end{array}
\right. .
\eqne
This also shows that for the ansatz of $\mu$ given in eq.\ (\ref{ansatzhor}), eq.\ (\ref{unitarityeqn2}) has no solution. For $i=2$, there is only one element in the Weyl group, therefore, we have proven Lemma \ref{lemma2}. For the case when $i\geq 3$, we need to check that only the trivial Weyl reflection can give a solution of eq.\ (\ref{unitarityeqn}) for the ansatz in eq.\ (\ref{ansatzhor}) and $n_1=\left[\tilde{\mu}^{i}\right] + 3$. To determine this, it is sufficient to consider the fundamental Weyl reflections. Consider first $w_{(j)}$ where $2 \leq j \leq i-2$. For those Weyl reflections one has
\eqnb
w_{(j)}\left(\tilde{\mu}+\rho_{\mf{h}'}-\left(\left[\tilde{\mu}^{i}\right]+2\right)\alpha^{(i)}\right) 
&=&
\tilde{\mu}+\rho_{\mf{h}'}-\left(\left[\tilde{\mu}^{i}\right]+2\right)\alpha^{(i)}-\left(\tilde{\mu}^{j}+1\right)\alpha^{(j)}.
\eqne
If one inserts this into eq.\ (\ref{unitarityeqn}), together with eq.\ (\ref{ansatzhor}), one finds that the sum is non-zero if
\eqnb
n_k
&=&
\left\{
\begin{array}{ll}
\left[\tilde{\mu}^{i}\right]+3-\left(\tilde{\mu}^{j}+1\right)\delta^{j}_k & 2 \leq k \leq i-1  \\
0                              & i \leq k \leq r_{\mf{g}}
\end{array}
\right.
\eqne
which is not possible as $n_{k+1}\leq n_{k}$ for all $k$. If we consider $w=w_{(i-1)}$ one will get
\eqnb
w_{(j)}\left(\tilde{\mu}+\rho_{\mf{h}'}-\left(\left[\tilde{\mu}^{i}\right]+2\right)\alpha^{(i)}\right) 
&=&
\tilde{\mu}+\rho_{\mf{h}'}-\left(\left[\tilde{\mu}^{i}\right]+2\right)\alpha^{(i)}-\left(\tilde{\mu}^{j}+\left[\tilde{\mu}^{i}\right]+3\right)\alpha^{(j)}.\no
\eqne
Inserting this into eq.\ (\ref{unitarityeqn}), and using eq.\ (\ref{ansatzhor}), implies that the sum is non-zero if
\eqnb
n_k
&=&
\left\{
\begin{array}{ll}
\left[\tilde{\mu}^{i}\right]+3 & 2 \leq k \leq i-2  \\
-\tilde{\mu}^{i-1}             & k = i-1 \\
0                              & i \leq k \leq r_{\mf{g}}
\end{array}
\right.
\eqne
which is not possible as $\tilde{\mu}^{i-1}\geq1$. Therefore, only the trivial Weyl reflection contributes. We have now proven Lemma \ref{lemma1}. $\Box$

\begin{lemma}
Let $\hat{\tilde{\mu}}$ be a strictly dominant highest weight where all components of $\tilde{\mu}$ are positive integers, $\tilde{\mu}^{r_{\mf{g}}}\neq0$ and $\tilde{k}\notin\Z_+$. Then there exists an antidominant highest weight $\hat\mu$ such that the state-space is non-unitary. 
\label{lemma4}
\end{lemma}

{\bf Proof:} With the assumptions of the lemma, the character can be written as
\eqnb
\chi^{\mf{h}'}_{\tilde{\mu}}\left(\theta,q\right)
&=&
\frac{\sum_{w\in W_{\mf{h}'}} \sign(w)e^{\ii\left(w\left(\tilde{\mu} + \rho_{\mf{h}'}\right)-\left(\tilde{\mu} + \rho_{\mf{h}'}\right),\theta\right)}}{\prod_{\alpha\in\Delta_c^+}\left(1-e^{-\ii\left(\alpha,\theta\right)}\right)\prod_{m=1}^\infty\left(1-q^m\right)^{r_\mf{g}-1}\prod_{\alpha\in\Delta_c}\left(1-q^m e^{\ii\left(\alpha,\theta\right)}\right)}+{\cal O}(q^N).\no
\eqne
Here $N=\left[\tilde{k}\right]-\left(\tilde{\mu},\theta'_{\mf{h}'}\right) + 3$ and we have, for convenience, excluded the factor involving the highest weight. This expression is clearly true in the limit $\tilde{k}\rightarrow \infty$ and validity of the expression set by $N$ follows from the Kac-Kazhdan determinant \cite{Sapovalov,Kac:1979fz}.

The signature function is
\eqnb
\Sigma^{\mf{h}'}_{\tilde{\mu}}\left(\theta,q\right)
&=&
\frac{\sum_{w\in W_{\mf{h}'}} \sign(w)e^{\ii\left(w\left(\tilde{\mu} + \rho_{\mf{h}'}\right)-\left(\tilde{\mu} + \rho_{\mf{h}'}\right),\theta\right)}}{\prod_{\alpha\in\Delta_c^+}\left(1-e^{-\ii\left(\alpha,\theta\right)}\right)\prod_{m=1}^\infty\left(1-q^m\right)^{r_\mf{g}-1}\prod_{\alpha\in\Delta_c}\left(1-q^m e^{\ii\left(\alpha,\theta\right)}\right)}
\no
&-&
2\sum_{m=0}q^{\left[\tilde{k}\right]-\left(\tilde{\mu},\theta'_{\mf{h}'}\right) + 2 + 2m}e^{\ii\left(\left[\tilde{k}\right]-\left(\tilde{\mu},\theta'_{\mf{h}'}\right) + 2 + 2m\right)\left(\theta'_{\mf{h}'},\theta\right)}\left(1-qe^{\ii\left(\theta'_{\mf{h}'},\theta\right)}\right)
\no
&\hspace{-7mm}\times&
\hspace{-7mm}\frac{\sum_{w\in W_{\mf{h}'}}\sign(w) e^{\ii\left(w\left(\left(\left[\tilde{k}\right]-\left(\tilde{\mu},\theta'_{\mf{h}'}\right) + 2 + 2m\right)\theta'_{\mf{h}'} + \tilde{\mu} + \rho_{\mf{h}'}\right)-\left(\left(\left[\tilde{k}\right]-\left(\tilde{\mu},\theta'_{\mf{h}'}\right) + 2 + 2m\right)\theta'_{\mf{h}'} + \tilde{\mu} + \rho_{\mf{h}'}\right),\theta\right)}}{\prod_{\alpha\in\Delta_c^+}\left(1-e^{-\ii\left(\alpha,\theta\right)}\right)\prod_{m=1}^\infty\left(1-q^m\right)^{r_\mf{g}-1}\prod_{\alpha\in\Delta_c}\left(1-q^m e^{\ii\left(\alpha,\theta\right)}\right)}
\no
&+&
{\cal O}(q^N),
\label{signatureaffineproblem}
\eqne
where $\theta'_{\mf{h}'}$ is the highest root of $\mf{h}'^{\C}$. This expression also follows from the Kac-Kazhdan formula. The second term arises from highest weight states w.r.t.\ the horizontal part of the algebra. These are of the form $\left(E_{-1}^{\theta'_{\mf{h}'}}\right)^{\left[\tilde{k}\right]-\sum_{l=2}^{r_\mf{g}}\tilde{\mu}^l+2}\left|\tilde{\mu},\tilde{k},0\right>$. 

Using the above expressions in the integral, eq.\ (\ref{chi-sigma}), and only keeping the terms of lowest order in $q$ gives
\eqnb
&&
\hspace*{-10mm}
2\sum_{n_1=0}^\infty e^{-\ii n_1\phi} \sum_{n_2=0}^{n_1}\ldots\sum_{n_{r_\mf{g}}=0}^{n_{r_\mf{g}-1}}\sum_{w\in W_{\mf{h}'}}\sign(w) \int d\theta e^{\ii\left(\mu + \tilde{\mu} + 2\rho_{\mf{h}'} + n_1\Lambda_{(2)} + \left(\left[\tilde{k}\right]-\left(\tilde{\mu},\theta'_{\mf{h}'}\right) + 2\right)\theta'_{\mf{h}'} - \sum_{l=2}^{r_\mf{g}}n_l\alpha^{(l)},\theta\right)}
\no
&\times&
e^{\ii\left(w\left(\left(\left[\tilde{k}\right]-\left(\tilde{\mu},\theta'_{\mf{h}'}\right) + 2\right)\theta'_{\mf{h}'} + \tilde{\mu} + \rho_{\mf{h}'}\right)-\left(\left(\left[\tilde{k}\right]-\left(\tilde{\mu},\theta'_{\mf{h}'}\right) + 2\right)\theta'_{\mf{h}'} + \tilde{\mu} + \rho_{\mf{h}'}\right),\theta\right)}=0.
\eqne
If this equation is not satisfied we have non-unitarity. The non-zero contribution to the integral over $\theta$ arises when 
\eqnb
\sum_{l=2}^{r_\mf{g}}n_l\alpha^{(l)}
&=&
\mu + \tilde{\mu} + 2\rho_{\mf{h}'} + n_1\Lambda_{(2)} + \left(\left[\tilde{k}\right]-\left(\tilde{\mu},\theta'_{\mf{h}'}\right) + 2\right)\left(\Lambda_{(2)}+\Lambda_{(r_{\mf{g}})}\right) 
\no
&+& w\left(\left(\left[\tilde{k}\right]-\left(\tilde{\mu},\theta'_{\mf{h}'}\right) + 2\right)\left(\Lambda_{(2)}+\Lambda_{(r_{\mf{g}})}\right) + \tilde{\mu} + \rho_{\mf{h}'}\right) 
\no
&-&
\left(\left(\left[\tilde{k}\right]-\left(\tilde{\mu},\theta'_{\mf{h}'}\right) + 2\right)\left(\Lambda_{(2)}+\Lambda_{(r_{\mf{g}})}\right) + \tilde{\mu} + \rho_{\mf{h}'}\right).
\label{affinecond:first}
\no
\eqne
Any non-trivial solution to the equations implies, as before, non-unitarity. In this equation $n_1$ is arbitrary and $n_{1}\geq n_{2}\geq \ldots \geq n_{r_{\mf{g}}}\geq 0$. We take as an ansatz that $n_{1}= n_{2}= \ldots = n_{r_{\mf{g}}}$ and determine whether this particular ansatz contributes to the integral. After this we will show that no other values of $n_{2}, \ldots , n_{r_{\mf{g}}}$ contribute for this value of $n_1$. We first find a solution for $w=1$ and then show that no other elements of the Weyl group contribute for the specific solution. 

Setting $w=1$ we determine the condition on $n_1$ for the affine weight $\hat{\mu}$ to be antidominant. Rewrite the previous equation using that $n_{1}= n_{2}= \ldots = n_{r_{\mf{g}}}$ gives
\eqnb
\hspace*{-5mm}
\mu
&=&
-\left\{
\tilde{\mu} + 2\rho_{\mf{h}'} + n_1\Lambda_{(2)} + \left(\left[\tilde{k}\right]-\left(\tilde{\mu},\theta'_{\mf{h}'}\right) + 2\right)\left(\Lambda_{(2)}+\Lambda_{(r_{\mf{g}})}\right)\right\} + n_1\sum_{l=2}^{r_\mf{g}}\alpha^{(l)}.
\label{affinecond}
\eqne
The requirement that the affine weight $\hat{\mu}$ is antidominant implies
\eqnb
k &<& \left(\mu,\theta'_{\mf{g}}\right)
\no
&=&
\sum_{l=1}^{r_\mf{g}}\mu^l
\no
&=&
\mu^1 + \left(\tilde{\mu},\theta'_{\mf{h}'}\right) + 2g^\vee_{\mf{h}'} + 2\left[k\right] + n_1,
\label{eq333}
\eqne 
where we have used that $r_\mf{g}=g^\vee_{\mf{h}'}$ and $\left[\tilde{k}\right] = -\left[k\right] - 2g^\vee_{\mf{h}'}-1$. 

Let us set
\eqnb
n_{1}
&=&
- \left[k\right] - \sum_{l=2}^{r_\mf{g}}\tilde{\mu}^l - 2g^\vee_{\mf{h}'} + 2 - a
\no
&=&
\left[\tilde{k}\right] - \sum_{l=2}^{r_\mf{g}}\tilde{\mu}^l + 3 - a
\no
&\geq&
3 - a,
\label{eq431}
\eqne
where $a\in \Z$. Inserting this into eq.\ (\ref{eq333}) yields
\eqnb
k-\left[k\right] &<& \mu^1 + 2 - a.
\eqne
Since $0 < k - \left[k\right] < 1$ and $\mu^1<-1$, the smallest possible value of $n_1$ arises when $a=0$. The minimum value of $n_1$ is therefore
\eqnb
n_1&=&\left[\tilde{k}\right] - \sum_{l=2}^{r_\mf{g}}\tilde{\mu}^l + 3
\no
&\geq&
3,
\label{eq432}
\eqne
and $\mu^1$ has to satisfy the inequality
\eqnb
k-\left[k\right] < \mu^1 + 2 < 1. 
\eqne 

We also have to check that the grade zero part of the weight is antidominant for this value of $n_1$. Consider first the $r_{\mf{g}}$ component of eq.\ (\ref{affinecond}),
\eqnb
\mu^{r_{\mf{g}}}
&=&
-\left(\tilde{\mu}^{r_{\mf{g}}} + 1\right).
\eqne
Using that $\tilde{\mu}^{r_{\mf{g}}}\geq1$ we find that the ansatz is consistent this far. It is straightforward to check the consistency of the ansatz by studying the other components of eq.\ (\ref{affinecond}). For this value of $n_1$ we have now that eq.\ (\ref{affinecond:first}) for $w=1$ can be written as
\eqnb
\left(\left[\tilde{k}\right] - \sum_{l=2}^{r_\mf{g}}\tilde{\mu}^l + 3\right)\sum_{l=2}^{r_\mf{g}}\alpha^{(l)}
&=&
\mu + \tilde{\mu} + 2\rho_{\mf{h}'} + \left(\left[\tilde{k}\right] - \sum_{l=2}^{r_\mf{g}}\tilde{\mu}^l + 3\right)\Lambda_{(2)} 
\no
&+&
 \left(\left[\tilde{k}\right]-\left(\tilde{\mu},\theta_{\mf{h}'}\right) + 2\right)\theta_{\mf{h}'}.
\eqne
Consider other values of $n_{2},\ldots,n_{r_{\mf{g}}}$. Assume first that $n_1=n_2$. By eq.\ (\ref{eq333}), we find that $n_1=n_{r_{\mf{g}}}$ and hence this is the same solution as previously considered. We now assume $n_1>n_2$. Inserting this into eq.\ (\ref{eq333}) implies
\eqnb
n_{r_{\mf{g}}}&>&\left[\tilde{k}\right] - \sum_{l=2}^{r_\mf{g}}\tilde{\mu}^l + 3,
\eqne
which shows that $n_{r_\mf{g}}$ is larger than $n_1$ thus inconsistent.

We furthermore need to show that no other Weyl reflection contributes to this sum. It is sufficient to show this for the simple Weyl reflections. The only element which is non-trivial is the one corresponding to the simple root $\alpha^{(r_\mf{g})}$,
\eqnb
w_{(r_\mf{g})}\left(\left(\left[\tilde{k}\right]-\left(\tilde{\mu},\tilde{\theta}\right) + 2\right)\left(\Lambda_{(2)}+\Lambda_{(r_{\mf{g}})}\right) + \tilde{\mu} + \rho_{\mf{h}'}\right)-
\no
\left(\left(\left[\tilde{k}\right]-\left(\tilde{\mu},\tilde{\theta}\right) + 2\right)\left(\Lambda_{(2)}+\Lambda_{(r_{\mf{g}})}\right) + \tilde{\mu} + \rho_{\mf{h}'}\right)
&=&
\no
\left(\left[\tilde{k}\right]-\sum_{l=2}^{r_\mf{g}}\tilde{\mu}^l + 3 + \tilde{\mu}^{r_\mf{g}}\right)\alpha^{(r_\mf{g})}.
\eqne 
As we have assumed that $\tilde{\mu}^{r_\mf{g}}\geq1$, the term corresponding to this Weyl reflection does not contribute. This concludes the proof of the lemma. $\Box$

Let us summarize the results for $\mf{g}^\C = A_{r_\mf{g}}$ in a theorem.
\begin{theorem}
Let $\hat{\tilde{\mu}}$ be a highest weight which satisfies $\tilde{\mu}^l>0$ for all $l$ and is strictly dominant. Furthermore, let $\tilde{\mu}^l\notin\Z_+$ for at least one value of $l$ or $\tilde{k}\notin\Z_+$. Then there exists at least one antidominant highest weight $\hat{\mu}$ such that the state-space is non-unitary.
\label{theorem2}
\end{theorem}
\noindent
{\bf Proof:} If there exist values $l$ such that $\tilde{\mu}^l\notin\Z_+$, then we can choose the smallest value. Use Lemma \ref{lemma1} and $\mu^j\geq1$ for $2\leq j\leq l-1$ or Lemma \ref{lemma2}, if $l=2$ which implies that the state-space is non-unitary. Therefore, the only case we need to consider is the one where the components of the horizontal weights are integers larger or equal to one and $\tilde{k}$ non-integer. For this case we can use Lemma \ref{lemma4} and $\mu^{r_{\mf{g}}}>0$, which implies that one can find an antidominant $\hat{\mu}$ such that the state-space is non-unitary. $\Box$

Note that integer values of $\tilde{\mu}^l$ and $\tilde{k}$ imply integer values of ${\mu}^l$ and ${k}$. The theorem gives almost the most general case within our assumption of antidominant highest weights $\hat\mu$. The cases that are not covered within Theorem \ref{theorem2} are when one or more components of $\mu$ are zero. Although we have made some progress in this direction, we have not been able to complete the proof for these cases.

Another case where one can reformulate Lemma \ref{lemma1} and Lemma \ref{lemma2} to hold for other coset constructions as well is $\mf{g} = \mf{su}(p,q)$. The reason is that one can consider the $\mf{su}(p)$ or the $\mf{su}(q)$ subalgebras independently at the grade zero level. This yields the models $\mf{su}(p,1)$ with subalgebra $\mf{su}(p)$ and $\mf{su}(q,1)$ with subalgebra $\mf{su}(q)$, respectively. The only difference is that one has to be careful to ensure that the weight $\hat{\mu}$ is antidominant. The difference in Lemma \ref{lemma1} and Lemma \ref{lemma2} is that the highest weight, apart from being strictly dominant, has to satisfy
\eqnb
\tilde{k}_1 + \tilde{k}_2 - 1 - \left[\tilde{\mu}^i\right] & > & \left(\tilde{\mu},\theta'_{\mf{h}'_1}\right) + \left(\tilde{\mu},\theta'_{\mf{h}'_2}\right) \no
\tilde{k}_1 + \tilde{k}_2 - \left[\tilde{\mu}^i\right] & > & \left(\tilde{\mu},\theta'_{\mf{h}'_1}\right) + \left(\tilde{\mu},\theta'_{\mf{h}'_2}\right),
\eqne
respectively.

We will in the next two sections, by studying embeddings of low rank algebra modules, find necessary conditions for more cases.


\sect{Necessary conditions from embedded submodules}

In this section we will consider submodules embedded in the module where the BRST charge acts in a simple way. These submodules will correspond to modules of $\mf{su}(2,1)$- or $\mf{sp}(4,\R)$-representations. To be able to consider these we need to study the representation theory of $\mf{su}(2)$ for strictly dominant highest weights in more detail, which is a special case of the previous section. Consider an $\mf{su}(2)$ subalgebra corresponding to the simple root $\alpha^{(i)}$, embedded in a larger algebra. The character minus the signature function is
\eqnb
\chi^{\mf{h}'}_{\tilde{\mu}}\left(\theta\right)-\Sigma^{\mf{h}'}_{\tilde{\mu}}\left(\theta\right) &=& 2\sum_{m=0}^\infty e^{\ii\left(\tilde\mu-\left(\left[\tilde{\mu}^{i}\right]+2m+2\right)\alpha^{(i)},\theta\right)},
\label{eq51}
\eqne 
where a factor proportional to $q$ has been suppressed. The character for the ghost sector is simple
\eqnb
\chi^{\mf{h}'}_{\tilde{\mu}}\left(\theta\right) 
&=&
e^{2\ii\left(\rho_{\mf{h}'},\theta\right)}\left(1-e^{-\ii\left(\alpha^{(i)},\theta\right)}\right)^2,
\label{eq51b}
\eqne 
where again the factor proportional to $q$ has been suppressed.

We will also need the expansion for an embedded $\mf{su}(2,1)$-module
\eqnb
&&
\hspace*{-10mm}
\frac{e^{\ii\left(\mu,\theta\right)}}{
\left(1-e^{-\ii(\alpha^{(i)},\theta)}\right)
\left(1-e^{-\ii\phi-\ii\left(\alpha,\theta\right)}\right)\left(1-e^{-\ii\phi-\ii(\alpha+\alpha^{(i)},\theta)}\right)} \no
&=&
\frac{e^{\ii\left(\mu,\theta\right)}}{\left(1-e^{-\ii(\alpha^{(i)},\theta)}\right)}\sum_{n_1=0}^\infty\sum_{n_2=0}^{n_1}e^{-\ii n_1\left(\phi+\left(\alpha,\theta\right)\right)-\ii n_2\left(\alpha^{(i)},\theta\right)},
\label{eq51c}
\eqne
where we have chosen a root $\alpha\in\Delta_+^n$ such that $\alpha+\alpha^{(i)}$ is a root but $\alpha-\alpha^{(i)}$ and $\alpha+2\alpha^{(i)}$ are not. Once again, a factor proportional to $q$ and $e^{\phi}$ corresponding to the highest weight has been suppressed. Eq.\ (\ref{eq51c}) follows from case 1 in Proposition \ref{characterexpansion}. Combining eq.\ (\ref{eq51c}) with the ghost contribution in eq.\ (\ref{eq51b}) we get $\chi^1_{\mu}\left(\phi,\theta\right)$ for this submodule
\eqnb
\chi^1_{\mu}(\phi,\theta) &=& e^{\ii\left(\mu+2\rho_{\mf{h}'},\theta\right)}\sum_{n_1=0}^\infty\sum_{n_2=0}^{n_1}e^{-\ii n_1\left(\phi+\left(\alpha,\theta\right)\right)-\ii n_2\left(\alpha^{(i)},\theta\right)}\left(1-e^{-\ii(\alpha^{(i)},\theta)}\right).
\label{eq52}
\eqne
Furthermore, we also need the expansion for an embedded $\mf{sp}(4,\R)$-module, which up to the highest weight factor involving $q$ and $e^{\phi}$, is
\eqnb
&&\hspace*{-10mm}
\frac{e^{\ii\left(\mu,\theta\right)}}{\left(1-e^{-\ii(\alpha^{(i)},\theta)}\right)\left(1-e^{-\ii\phi-\ii\left(\alpha,\theta\right)}\right)\left(1-e^{-\ii\phi-\ii(\alpha+\alpha^{(i)},\theta)}\right)\left(1-e^{-\ii\phi-\ii(\alpha+2\alpha^{(i)},\theta)}\right)} 
\no
&=&
\frac{e^{\ii\left(\mu,\theta\right)}}{\left(1-e^{-\ii(\alpha^{(i)},\theta)}\right)}\sum_{n_1=0}^\infty\sum_{n_2=0}^\infty\sum_{n_3=0}^{2n_1}e^{-\ii n_1\left(\phi+\left(\alpha,\theta\right)\right)- 2\ii n_2\left(\phi+\left(\alpha + \alpha^{(i)},\theta\right)\right) - \ii n_3\left(\alpha^{(i)},\theta\right)},
\label{eq53a}
\eqne
where we have chosen a root $\alpha\in\Delta_+^n$ such that $\alpha+\alpha^{(i)}$ and $\alpha+2\alpha^{(i)}$ are roots but $\alpha-\alpha^{(i)}$ and $\alpha+3\alpha^{(i)}$ are not. One can prove this identity by expanding the denominator on the right-hand side or apply techniques used in the proof of Proposition \ref{characterexpansion}. Combining eq.\ (\ref{eq53a}) with the ghost contribution in eq.\ (\ref{eq51b}) we get $\chi^1_{\mu}(\phi,\theta)$ for this submodule
\eqnb
\chi^1_{\mu}(\phi,\theta) &=& e^{\ii\left(\mu+2\rho_{\mf{h}'},\theta\right)}\left(1-e^{-\ii(\alpha^{(i)},\theta)}\right)
\no
&\times&
\sum_{n_1=0}^\infty\sum_{n_2=0}^\infty\sum_{n_3=0}^{2n_1}e^{-\ii n_1\left(\phi+\left(\alpha,\theta\right)\right)- 2\ii n_2\left(\phi+\left(\alpha + \alpha^{(i)},\theta\right)\right) - \ii n_3\left(\alpha^{(i)},\theta\right)}.
\label{eq53}
\eqne

\begin{theorem}
Let $\mf{g}=\mf{su}(r_{\mf{g}},1)$ and $\hat{\tilde{\mu}}$ be a strictly dominant weight. If $\tilde{\mu}^2\notin \Z_+$ or $\tilde{\mu}^{r_{\mf{g}}}\notin \Z_+$ then there exists at least one  antidominant weight $\hat\mu$ such that the state-space is non-unitary. 
\label{theorem3}
\end{theorem}
{\bf Proof:}
Consider first the simple root $\alpha^{(2)}$. This case is proven in Lemma \ref{lemma2}, but let us use another technique to prove it. We consider a subset of creation operators $E_0^{-\alpha^{(1)}}$, $E_0^{-\alpha^{(2)}}$ and $E_0^{-\left(\alpha^{(1)}+\alpha^{(2)}\right)}$. These operators create an $\mf{su}(2,1)$ submodule. Consider this submodule together with the three submodules created by $\tilde{E}_0^{-\alpha^{(2)}}$, $c^{-\alpha^{(2)}}_0$ and $b^{-\alpha^{(2)}}_0$. When one acts with the BRST charge on this subspace of the full state-space many terms in the charge act trivially. A reason for this is that $\alpha^{(1)}+\alpha^{(2)}-\alpha^{(k)}$ is not a root for $k\neq 2$. Therefore, the only part of the BRST charge which acts non-trivially on this space is the one containing the operators corresponding to the simple root $\alpha^{(2)}$. Thus, one can study this subspace to get necessary conditions. Let $\tilde{\mu}^{2}$ be non-integer, then eq.\ (\ref{chi-sigma}) for this subspace is
\eqnb
\sum_{m}^\infty\sum_{n_1}^\infty\sum_{n_2}^{n_1} e^{-\ii n_1\phi} \int d\theta \left(1-e^{-\ii\left(\alpha^{(2)},\theta\right)}\right)e^{\ii\left(\mu + \tilde\mu + 2\rho_{\mf{h}'} + n_1\Lambda_{(2)}- \left(\left[\tilde\mu^{2}\right] + 2 + 2m + n_2\right)\alpha^{(2)},\theta\right)}=0,
\label{eq54}
\eqne
where eqs.\ (\ref{eq51}) and (\ref{eq52}) have been used with $i=2$, $\alpha=\alpha^{(1)}$ and $-\left(\alpha,\theta\right) = \left(\Lambda_{(2)},\theta\right)$. If this integral has a non-trivial value, then we have non-unitarity. The integral is non-zero if
\eqnb
\mu + \tilde\mu + 2\rho_{\mf{h}'} + n_1\Lambda_{(2)}- \left(\left[\tilde\mu^{2}\right] + 2 + 2m + n_2\right)\alpha^{(2)} &=& 0.
\eqne 
A solution to this equation is
\eqnb
n_1     &=& \left[\tilde{\mu}^{2}\right] + 3 \no
\mu^{l} &=& - \tilde{\mu}^l - 2 + \left(\left[\tilde{\mu}^{2}\right] + 1\right)\delta_{2,l} - \left(\left[\tilde{\mu}^{2}\right] + 2\right)\delta_{3,l}.
\eqne
This solution yields $n_2 = m = 0$. Here one can see that the second term in eq.\ (\ref{eq54}) does not contribute. One now has to check that the highest weight is antidominant. This is the same analysis as for the proof of Lemma \ref{lemma2} in the previous section. The analysis shows that the highest weight is indeed antidominant if the weight $\hat{\tilde{\mu}}$ is strictly dominant.

The creation operators $E_{-1}^{\theta'_{\mf{g}}}$, $E_{0}^{-\alpha^{(r_{\mf{g}})}}$ and $E_{-1}^{\theta'_{\mf{g}}-\alpha^{(r_{\mf{g}})}}$, where $\theta'_{\mf{h}'}$ is the highest root of $\mf{g}^{\C}$, will create an $\mf{su}(2,1)$-submodule. We study this submodule together with the submodule generated by $\tilde{E}_0^{-\alpha^{(r_{\mf{g}})}}$, $c^{-\alpha^{(r_{\mf{g}})}}_0$ and $b^{-\alpha^{(r_{\mf{g}})}}_0$. As $\theta'_{\mf{g}} - \alpha^{(r_{\mf{g}})} + \alpha^{(k)}$ is not a root for $k\neq r_{\mf{g}}$, all operators in the BRST operator with grade zero except the one corresponding to the simple root $\alpha^{(r_{\mf{g}})}$ act trivially. Since $E_{0}^{\theta'_{\mf{g}}-\theta'_{\mf{h}'}}$ is an annihilation operator and $\theta'_{\mf{g}}-\alpha^{(r_{\mf{g}})}-\theta'_{\mf{h}'}$ is not a root, the operator corresponding to the simple non-horizontal root acts trivially. Therefore, the terms that act non-trivially on the subspace correspond to a BRST charge for the $\mf{su}(2)$-subalgebra generated by the operators corresponding to the simple root $\alpha^{(r_{\mf{g}})}$. One can now proceed in the same way as for the $\mf{su}(2)$-submodule corresponding to the simple root $\alpha^{(2)}$. Choosing a non-integer weight $\tilde\mu^{2}$, eq.\ (\ref{chi-sigma}) yields
\eqnb
\sum_{m}^\infty\sum_{n_1}^\infty\sum_{n_2}^{n_1} e^{-\ii n_1\phi} \int d\theta \left(1-e^{-\ii\left(\alpha^{(r_{\mf{g}})},\theta\right)}\right)e^{\ii\left(\mu + \tilde\mu + 2\rho_{\mf{h}'} + n_1\Lambda_{(r_{\mf{g}})}- \left(\left[\tilde\mu^{r_{\mf{g}}}\right] + 2 + 2m + n_2\right)\alpha^{(r_{\mf{g}})},\theta\right)}=0,\no
\label{eq57}
\eqne
where eqs.\ (\ref{eq51}) and (\ref{eq52}) have been used\footnote{Observe that the character is not precisely the one given above, it should be multiplied by a factor $q^{n_1}$. As this factor is unimportant we suppress it.} together with $i=r_\mf{g}$, $\alpha=\theta'_{\mf{g}}$ and $-\left(\theta'_{\mf{g}},\theta\right)=\left(\Lambda_{(\mf{g})},\theta\right)$. If this equation is not satisfied, we have non-unitarity. The integral has a non-zero value if
\eqnb
\mu + \tilde\mu + 2\rho_{\mf{h}'} + n_1\Lambda_{(r_{\mf{g}})}- \left(\left[\tilde\mu^{r_{\mf{g}}}\right] + 2 + 2m + n_2\right)\alpha^{(r_{\mf{g}})} &=& 0
\eqne 
A solution to these equations is
\eqnb
n_1     &=& \left[\tilde{\mu}^{r_{\mf{g}}}\right] + 3 \no
\mu^{l} &=& - \tilde{\mu}^l - 2 + \left(\left[\tilde{\mu}^{r_{\mf{g}}}\right] + 1\right)\delta_{r_{\mf{g}},l} - \left(\left[\tilde{\mu}^{r_{\mf{g}}}\right] + 2\right)\delta_{r_{\mf{g}}-1,l}.
\label{eq59}
\eqne
This solution implies that $n_2 = m = 0$, which shows that the second term in eq.\ (\ref{eq57}) does not contribute. We now need to check that the solution $\hat\mu$ is antidominant. Inserting the solution given in eq.\ (\ref{eq59}) into $k<\left(\mu,\theta'_{\mf{g}}\right)$ yields
\eqnb
\tilde{k} + 1 + \mu^1 &>& \left(\tilde\mu,\theta'_\mf{h'}\right)
\eqne
which confirms that the highest weight is antidominant if $\hat{\tilde{\mu}}$ is a strictly dominant highest weight. Combining the two above submodules proves the theorem. $\Box$

Let us consider other examples where we can prove that a few of the components of the highest weight have to be integers. But to be able to do this we have to be more restrictive on which types of highest weights we can allow. We summarize the results in a theorem.

\begin{theorem}
Let $\tilde{\mu}^j>-1$ for $j=2,\ldots,r_\mf{g}$. If the below conditions are satisfied then there exists at least one antidominant weight $\hat{\mu}$ such that the state-space is non-unitary. 
\begin{enumerate}
\item[i)] {$\mf{g}=\mf{su}(p,q)$: $\tilde{\mu}^i\notin\Z_+$ for $i=2,p,p+1$ or $p+q-1$; $\tilde{k}_1 + \tilde{k}_2 - \left[\tilde{\mu}^{i}\right] - 1 > \left(\tilde\mu,{\theta'}^\vee_{\mf{h}'_1}\right) + \left(\tilde\mu,{\theta'}^\vee_{\mf{h}'_2}\right)$}

\item[ii)] {$\mf{g}=\mf{sp}(2r_\mf{g},\R)$: $\tilde{\mu}^i\notin\Z_+$ for $i=2$ or $r_\mf{g}$; $\tilde{\kappa} - 2g^\vee_{\mf{h}'}>2\left(\tilde{\mu},{\theta'}^\vee_{\mf{h}'}\right)$.}

\item[iii)] {$\mf{g}=\mf{so}(2r_\mf{g}-1,2)$: $\tilde{\mu}^2\notin\Z_+$; $\tilde{k} - 3 - \tilde{\mu}^i  > \left(\tilde{\mu},{\theta'}^\vee_{\mf{h}'}\right)$}

\item[iv)] {$\mf{g}=\mf{so}(2r_\mf{g}-2,2)$: $\tilde{\mu}^2\notin\Z_+$; $\tilde{k} - 3 - \tilde{\mu}^i  > \left(\tilde{\mu},{\theta'}^\vee_{\mf{h}'}\right)$}

\item[v)] {$\mf{g}=\mf{so}^*(2r_\mf{g})$: $\tilde{\mu}^i\notin\Z_+$ for $i=3,r_\mf{g}-1$; $\tilde{k} + 3 - 2g^\vee_{\mf{h}'} - \sum_{l=3}^{r_{\mf{g}}-1}\tilde{\mu}^{i} - \left[\tilde{\mu}^3\right] > \left(\tilde{\mu},{\theta'}^\vee_{\mf{h}'}\right)$}

\item[vi)] {$\mf{g}=E_{6|-14}$: $\tilde{\mu}^2\notin\Z_+$; $\tilde{k} - 9 - \tilde{\mu}^{2} - \tilde{\mu}^{3} - \tilde{\mu}^{6} - \left[\tilde{\mu}^2\right] > \left(\tilde{\mu},{\theta'}^\vee_{\mf{h}'}\right)$}

\item[vii)] {$\mf{g}=E_{7|-25}$: $\tilde{\mu}^i\notin\Z_+$ for $i=2,6$; $\tilde{k} - 13 - \sum_{l=2}^{6}\tilde{\mu}^{i} - \left[\tilde{\mu}^i\right] > \left(\tilde{\mu},{\theta'}^\vee_{\mf{h}'}\right)$}
\end{enumerate}
\label{theorem4}
\end{theorem}
{\bf Proof:}
Let us study the values $i$ where $\alpha^{(1)}+\alpha^{(i)}\in \Delta^+$. This is the case for $i=2$ for $\mf{g}=\mf{su}(p,q)$, $\mf{sp}(2r_\mf{g})$, $\mf{so}(2r_\mf{g}-1,2)$, $\mf{so}(2r_\mf{g}-2,2)$, $E_{6|-14}$ and $E_{7|-25}$, $i=3$ for $\mf{g}=\mf{so}^*(2r_\mf{g})$ and $i=p+1$ for $\mf{g}=\mf{su}(p,q)$. Consider first the case where $\alpha^{(i)}$ is a long root. A subset of creation operators $E_0^{-\alpha^{(1)}}$, $E_0^{-\alpha^{(i)}}$ and $E_0^{-\left(\alpha^{(1)}+\alpha^{(i)}\right)}$ generate an $\mf{su}(2,1)$-submodule. Adding to this the submodule generated by $\tilde{E}_0^{-\alpha^{(i)}}$, $c^{-\alpha^{(i)}}_0$ and $b^{-\alpha^{(i)}}_0$ produces a state-space where the BRST charge acts as a simple $\mf{su}(2)$ BRST charge corresponding to the simple root $\alpha^{(i)}$. This follows since $\alpha^{(1)}+\alpha^{(i)}-\alpha^{(k)}$ is not a root for $k\neq i$. Inserting the character and signature functions for each sector, given by eqs.\ (\ref{eq51}) and (\ref{eq52}), into the integral given by eq. (\ref{chi-sigma}) yields an integral which is non-zero only if
\eqnb
\mu + \tilde\mu + 2\rho_{\mf{h}'} + n_1\left(\Lambda_{(i)}+a_{l,i}\Lambda_{(l)}\right)- \left(\left[\tilde\mu^{i}\right] + 2 + 2m + n_2\right)\alpha^{(i)} &=& 0,
\label{eq511}
\eqne
where $a_{l,i}$ is only non-zero for $\mf{g}=\mf{su}(p,q)$,
\eqnb
a_{l,i}
&=&
\left\{
\begin{array}{lcl}
1 && i=2;\,l=p+1 \\
1 && i=p+1;\,l=2 \\
\end{array}
\right. 
.
\eqne
A solution to these equations is
\eqnb
n_1     &=& \left[\tilde{\mu}^{i}\right] + 3 \no
\mu^{l} &=& - \tilde{\mu}^l - 2   -\left(\left[\tilde{\mu}^{i}\right]+3\right)\delta_{2,l} +\left(\alpha^{(i)},\alpha^{(l)^\vee}\right)\left(\left[\tilde{\mu}^{i}\right]+2\right) - a_{l,i}\left(\left[\tilde{\mu}^{i}\right]+3\right).
\eqne

Let us now study the case when the simple root $\alpha^{(i)}$ is short. There is only one such case and it is $i=2$ and $\mf{g}=\mf{sp}(2r_{\mf{g}},\R)$. When the simple root is short one has one additional creation operator, $E^{-\alpha^{(1)}-2\alpha^{(i)}}$. The submodule generated is then a module corresponding to an $\mf{sp}(4,\R)$ representation. The other sectors are the same as those studied in the previous paragraph and the BRST charge acts as an $\mf{su}(2)$ BRST charge. Inserting the character and signature function, given by eqs.\ (\ref{eq51}) and (\ref{eq53}), into eq.\ (\ref{chi-sigma}) yields
\eqnb
&&
\hspace*{-10mm}
2\sum_{n_1,n_2=0}^\infty\sum_{n_3=0}^{2n_1} e^{-\ii\left(n_1+2n_2\right)\phi}\int d\theta \left(1-e^{-\ii\left(\alpha^{(i)},\theta\right)}\right)
\no
&\times&
e^{\ii\left(\mu + \tilde{\mu} + 2\rho_{\mf{h}'} + 2n_1\Lambda_{(2)} + 2n_2\Lambda_{(3)} - \left(\left[\tilde\mu^{i}\right] + 2 + 2m + n_3\right)\alpha^{(2)},\theta\right)}=0.
\label{eq515}
\eqne
We get a non-zero result for the integral above when
\eqnb
n_1     &=& \frac{1}{2}\left(\left[\tilde{\mu}^{2}\right] + 3 + a\right)\no
\mu^{2} &=& - 1 - \left(\tilde{\mu}^{i}-\left[\tilde{\mu}^{i}\right]\right) - a\no
\mu^{l} &=& - \tilde{\mu}^l - 2 +\left(\alpha^{(2)},\alpha^{(l)^\vee}\right)\left(\left[\tilde{\mu}^{2}\right]+2\right) \phantom{1234} l\neq 2
\label{eq516}
\eqne
which yields that $n_2 = n_3 = m = 0$. Here $a$ is equal to zero or one if $\left[\tilde{\mu}^2\right]$ is an odd or even integer, respectively.

We now need to check that the solution we have for weight $\hat{\mu}$ is antidominant. We need to study this case by case as the relation between the highest weights are different in the different algebras. Inserting the solutions above in the inequality $k<\left(\mu,\theta'_{\mf{g}}\right)$ yields
$$
\begin{array}{rl}
\mf{su}(p,q): & \tilde{k}_1 + \tilde{k}_2 + \mu^{1} - \left[\tilde{\mu}^{i}\right] > \left(\tilde\mu,\theta'_{\mf{h}'_1}\right) + \left(\tilde\mu,\theta'_{\mf{h}'_2}\right) \\

\mf{sp}(2r_\mf{g},\R): &\tilde{\kappa} + 2 + 2\mu^1 - 2g^\vee_{\mf{h}'}>2\left(\tilde{\mu},{\theta'}^\vee_{\mf{h}'}\right) \\

\mf{so}(2r_\mf{g}-1,2):& \tilde{k} - 2 + \mu^1 - \tilde{\mu}^2  > \left(\tilde{\mu},{\theta'}^\vee_{\mf{h}'}\right)\\

\mf{so}(2r_\mf{g}-2,2):& \tilde{k} - 2 + \mu^1 - \tilde{\mu}^2  > \left(\tilde{\mu},{\theta'}^\vee_{\mf{h}'}\right)\\

\mf{so}^*(2r_\mf{g}):& \tilde{k} + 4 - 2g^\vee_{\mf{h}'} + \mu^1 - \sum_{l=3}^{r_{\mf{g}}-1}\tilde{\mu}^{i} - \left[\tilde{\mu}^3\right] > \left(\tilde{\mu},\theta'_{\mf{h}'}\right)\\

E_{6|-14}:& \tilde{k} - 8 + \mu^1 - \tilde{\mu}^{2} - \tilde{\mu}^{3} - \tilde{\mu}^{6} - \left[\tilde{\mu}^2\right] > \left(\tilde{\mu},\theta'_{\mf{h}'}\right)\\

E_{7|-25}:& \tilde{k} - 12 + \mu^1 - \sum_{l=2}^{6}\tilde{\mu}^{i} - \left[\tilde{\mu}^2\right] > \left(\tilde{\mu},\theta'_{\mf{h}'}\right)\\ 
\end{array}
$$
As $\mu^1<-1$ one can choose a highest weight $\hat\mu$ which is antidominant as long as the conditions of the theorem are satisfied. 

We consider now the values of $i$ for which $\theta'_{\mf{g}}-\alpha^{(i)}\in\Delta^+$. The values for which this is true are $i=p$ and $i=p+q-1$ for $\mf{su}(p,q)$, $i=r_\mf{g}$ for $\mf{sp}(2r_{\mf{g}},\R)$, $i=r_\mf{g}-1$ for $\mf{so}^*(2r_{\mf{g}})$ and $i=6$ for $E_{7|-25}$. For the other algebras the simple root is the same as previously studied and will not yield anything new. 

Let us first study the case when $\alpha^{(i)}$ is a long root. Consider the subset of creation operators, $E_{-1}^{\theta'_{\mf{g}}}$, $E_0^{-\alpha^{(i)}}$ and $E_{-1}^{\theta'_{\mf{g}}-\alpha^{(i)}}$. They give an $\mf{su}(2,1)$-submodule. Considering this submodule together with the submodule generated by $\tilde{E}_0^{-\alpha^{(i)}}$, $c^{-\alpha^{(i)}}_0$ and $b^{-\alpha^{(i)}}_0$ produces a state-space where the BRST charge acts as an $\mf{su}(2)$ BRST charge corresponds to the simple root $\alpha^{(i)}$. This follows since $\theta'_{\mf{g}}-\alpha^{(i)}+\alpha^{(k)}$ is not a root for $k\neq i$; and $\theta'_{\mf{g}}-\alpha$ is either not a root or a positive root for $\alpha\in \Delta^{+}_{\mf{h}'}$. Inserting eqs.\ (\ref{eq51}) and (\ref{eq52}) into eq.\ (\ref{chi-sigma}) we find an equation which is the same as eq.\ (\ref{eq511}), but with $a_{i,l}\rightarrow \tilde{a}_{i,l}$. This equation is only non-zero for $\mf{g}=\mf{su}(p,q)$ and 
\eqnb
\tilde{a}_{i,l}
&=&
\left\{
\begin{array}{lcl}
1 && i=p;\,l=p+q-1 \\
1 && i=p+q-1;\,l=p \\
\end{array}
\right. 
.
\eqne
A solution which gives a non-zero value for the integral is
\eqnb
n_1     &=& \left[\tilde{\mu}^{2}\right] + 3 \no
\mu^{i} &=& - 1 - \left(\tilde{\mu}^{i}-\left[\tilde{\mu}^{i}\right]\right) \no
\mu^{l} &=& - \tilde{\mu}^l - 2 +\left(\alpha^{(i)},\alpha^{(l)^\vee}\right)\left(\left[\tilde{\mu}^{i}\right]+2\right) - \tilde{a}_{l,i}\left(\left[\tilde{\mu}^{i}\right]+3\right) \phantom{1234} l\neq i
\eqne
This solution implies that $n_2 = m = 0$. 

Consider now the case when $\alpha^{(i)}$ is a short root. The only case here is $i=r_{\mf{g}}$ where $\mf{g}=\mf{sp}(2r_{\mf{g}},\R)$. The steps in the proof here are the same as for the other submodule where the simple root is long. The only difference is that one changes $\left(\Lambda_{(2)},\Lambda_{(3)},\alpha^{(2)}\right)\rightarrow\left(\Lambda_{(r_\mf{g})},\Lambda_{(r_\mf{g}-1)},\alpha^{(r_\mf{g})}\right)$ in eqs.\ (\ref{eq515}) and the solution (\ref{eq516}).

We need to check that the weight $\hat{\mu}$ is antidominant. Again this is a case by case study. Inserting the solution above in the inequality $k<\left(\mu,{\theta'}^{\vee}_{\mf{g}}\right)$ yields
$$
\begin{array}{rl}
\mf{su}(p,q): & \tilde{k}_1 + \tilde{k}_2 + \mu^{1} - \left[\tilde{\mu}^{i}\right] - 1 > \left(\tilde\mu,\theta'_{\mf{h}'_1}\right) + \left(\tilde\mu,{\theta'}^{\vee}_{\mf{h}'_2}\right) \\

\mf{sp}(2r_\mf{g},\R): & \tilde{\kappa} + 2 + 2\mu^1 - 2g^\vee_{\mf{h}'}>2\left(\tilde{\mu},{\theta'}^\vee_{\mf{h}'}\right) \\

\mf{so}^*(2r_\mf{g}):& \tilde{k} + 4 - 2g^\vee_{\mf{h}'} + \mu^1 - \sum_{l=3}^{r_{\mf{g}}-1}\tilde{\mu}^{i} - \left[\tilde{\mu}^{r_\mf{g}-1}\right] > \left(\tilde{\mu},{\theta'}^{\vee}_{\mf{h}'}\right)\\

E_{7|-25}:& \tilde{k} - 12 + \mu^1 - \sum_{l=2}^{6}\tilde{\mu}^{i} - \left[\tilde{\mu}^{6}\right] > \left(\tilde{\mu},{\theta'}^{\vee}_{\mf{h}'}\right)\\ 
\end{array}
$$
As $\mu^1<-1$ one can choose a highest weight $\hat\mu$ which is antidominant as long as the conditions of the theorem are satisfied. $\Box$


\sect{Further necessary conditions}
Let us summarize what we have been able to show so-far. We have shown, quite generally, necessary conditions for $\mf{g}=\mf{su}(r_{\mf{g}},1)$, due to the simplicity of root system in this case, and for the grade zero part of $\mf{g}=\mf{su}(p,q)$. Furthermore, we have been able to show that certain components of the highest weight $\hat{\mu}$ have to be integers by studying embedded submodules in the state-space. For algebras other than $\mf{g}=\mf{su}(r_{\mf{g}},1)$, or other components of the highest weight, we have not yet established any necessary conditions.

In this section we will take a slightly different route to give us further results on the representations for the auxiliary sector which gives non-unitarity. We will here consider the different cases in Proposition \ref{characterexpansion}. 
We use eq.~(\ref{chi1:hori}) together with the expansions given in Proposition  \ref{characterexpansion}  and insert them into eq.\ (\ref{Integration:horizontal}). This gives the following equation
\eqnb
\sum_{m=0}^{\infty}\sum_{n_i\in\Z_+}\sum_{w\in W_{\mf{h}'}}\sign(w) N_{n_i\alpha^{(i)}}e^{-\ii m\phi}\int d\theta e^{\ii \left(\theta,\mu + \tilde\mu + \rho_{\mf{h}'} + w(m\Lambda_{(2)}+\rho_{\mf{h}'}) - n_i\alpha^{(i)}\right)}=0
.
\label{Cond1.horizunitarity}
\eqne

We examine this equation to determine for which weights this is not satisfied. Assume that $\left(\mu+\tilde{\mu}+2\rho_{\mf{h}'} + m \Lambda_{(2)} - (m+1)\alpha^{(2)},\Lambda_{(2)}\right) < 0$, which will be checked in all examples later on. This restricts the elements of the Weyl group for which the integral may be non-zero. The restriction is that the simplest word for the Weyl element cannot involve $w_2$. Denote by $W''$ this set of Weyl transformations. Then 
\eqnb
\sum_{m=0}^{\infty}\sum_{n_i\in\Z_+}\sum_{w\in W''}\sign(w) N_{n_i\alpha^{(i)}}e^{-\ii m\phi}\int d\theta e^{ \ii \left(\theta,\mu + \tilde\mu + \rho_{\mf{h}'} + m\Lambda_{(2)} + w(\rho_{\mf{h}'}) - n_i\alpha^{(i)}\right)}=0
.
\label{Cond2.horizunitarity}
\eqne
We will in the following analyze this equation for components $\tilde\mu^i$ close to the non-compact simple root i.e.\ values of $i$ closest to one. The resulting conditions, although valid generally, are especially interesting for algebras of low rank, as then the conditions will be complete or almost complete necessary conditions for the horizontal components. We will furthermore restrict our analysis to weights satisfying $\tilde{\mu}^i>-1$.

\subsection{Conditions on $\tilde{\mu}^3$}
Let us investigate conditions on $\tilde{\mu}^3$ coming from unitarity. By Theorem \ref{theorem3} and Theorem \ref{theorem4} we know that $\tilde{\mu}^2$ is required to be an integer. This we can therefore assume henceforth. Before considering the conditions on $\tilde{\mu}^{3}$ we have to consider a few states in a non-unitary representation of $\mf{su}(3)$ and $\mf{so}(5)$, where $\tilde{\mu}^{2}\in\Z_+$ and $\tilde{\mu}^{3}\notin\Z_+$. These states are
\eqnb
\left|\phi\right> &=& \left(E^{-\alpha^{(2)}}\right)^{\tilde{\mu}+n_3}\left(E^{-\alpha^{(3)}}\right)^{n_3}\left|\tilde{\mu},\tilde{k},0\right>,
\eqne
which all have degeneration one. The norms of the states are for both algebras
\eqnb
\left<\phi\left|\phi\right>\right. &=& \left(\left(\tilde{\mu}^{2} + n_3\right)!\right)^2 n_2! \prod_{j=0}^{n_3-1}\left(\tilde{\mu}^3-j\right)
\no
&=&
\left\{
\begin{array}{ll}
\left|K\right|^2 & n_3\leq\left[\tilde{\mu}^{3}\right] + 1 \\
\left|K\right|^2(-)^{n_3 - \left[\tilde{\mu}^3\right] - 1} & n_3\geq\left[\tilde{\mu}^{3}\right] + 1,
\end{array}
\right.
\eqne
where $K$ is unimportant in the following. 

The equations following from the trivial Weyl reflection which gives non-zero contributions to the integral in eq.~(\ref{Cond2.horizunitarity}) is\footnote{We have here defined $\epsilon^i \equiv - 1 - \mu^i >0$.}
\eqnb
\tilde\mu^2 - \epsilon^2 + 1 + m &=& 2n_2 - n_3 \no
\tilde\mu^3 - \epsilon^3 + 1     &=& n_2\left(\alpha^{(2)},\alpha^{(3)^\vee}\right) + 2n_3 \no
\tilde\mu^l - \epsilon^l + 1     &=& n_2\left(\alpha^{(2)},\alpha^{(l)^\vee}\right) + n_3\left(\alpha^{(3)},\alpha^{(l)^\vee}\right),\phantom{emty}i = 4,\ldots,r_{\mf{g}}
\label{cond:rank3}
\eqne
where 
\eqnb
\left(\alpha^{(2)},\alpha^{(3)^\vee}\right) &=& 
\left\{
\begin{array}{cc}
-2 & \mf{g}^\C = B_3 \\
-1 & \mathrm{Otherwise}
\end{array}
\right.
\\
\left(\alpha^{(2)},\alpha^{(4)^\vee}\right) &=& 
\left\{
\begin{array}{cc}
-1 & \mf{g}^\C = D_4 \\
0 & \mathrm{Otherwise}
\end{array}
\right.
\\
\left(\alpha^{(3)},\alpha^{(4)^\vee}\right) &=& 
\left\{
\begin{array}{cc}
-2 & \mf{g}^\C = B_4 \\
0 & r_\mf{g} = 3 \\
-1 & \mathrm{Otherwise}
\end{array}
\right.
\\
\left(\alpha^{(3)},\alpha^{(5)^\vee}\right) &=& 
\left\{
\begin{array}{cc}
-1 & \mf{g}^\C = D_5 \\
0 & \mathrm{Otherwise}
\end{array}
\right.
.
\eqne
We fix 
\eqnb
m &=& \tilde\mu^2 + \left[\tilde\mu^3\right] + 2 \no
\epsilon^2 &=& 1 \no 
\epsilon^3 &=& -\left(\tilde\mu^2+2 +\left[\tilde\mu^3\right]\right)\left(\alpha^{(2)},\alpha^{(3)^\vee}\right) + \tilde\mu^3 - 3 - 2\left[\tilde\mu^3\right] \no
\epsilon^l &=& \tilde{\mu}^l + 1 - \left(\left[\tilde\mu^3\right] + 2 + \tilde\mu^2\right)\left(\alpha^{(2)},\alpha^{(l)^\vee}\right) - \left(\left[\tilde\mu^3\right] + 2\right)\left(\alpha^{(3)},\alpha^{(l)^\vee}\right),\phantom{emty}i = 4,\ldots,r_{\mf{g}}.
\label{eq610}
\no
\eqne
This corresponds, for the horizontal part, to an antidominant weight if $\tilde\mu^2 > 0$ for all cases except $\mf{g}^{\C}=B_3$. For $\mf{g}^{\C}=B_3$ we do not get any restrictions on $\tilde\mu^2$. Furthermore, we have an antidominant highest weight with $k<\left(\mu,{\theta'_{\mf{g}}}^\vee\right)$ if the following conditions are satisfied
\eqnb
\begin{array}{ll}
\mf{su}(r_{\mf{g}},1): & \tilde{k} + 2 + \mu^1 + \left(\left[\tilde{\mu}^{3}\right] + 2\right)\delta^{3}_{r_\mf{g}} > \left(\tilde{\mu},{\theta'_{\mf{h}'}}^\vee\right) \\
\mf{so}(2r_{\mf{g}}-1,2): & \tilde{k} + \mu^1 - \tilde{\mu}^2 > \left(\tilde{\mu},{\theta'_{\mf{h}'}}^\vee\right) \\
\mf{so}(2r_{\mf{g}}-2,2): & \tilde{k} + \mu^1 - \tilde{\mu}^2 > \left(\tilde{\mu},{\theta'_{\mf{h}'}}^\vee\right) \\
\end{array}
\nonumber
\eqne
Inserting the ansatz, eq.\ (\ref{eq610}), into eq.\ (\ref{cond:rank3}) implies 
\eqnb
n_2 &=& \tilde\mu^2 + \left[\tilde\mu^3\right] + 2 \no
n_3 &=& \left[\tilde\mu^3\right] + 2,
\eqne
and $n_l=0$ for $l=4,\ldots,r_\mf{g}$. We may again conclude that the only other Weyl transformations that contributes is $w_3$. Studying the equations shows that the solution is $n'_2 = \left[\tilde\mu^3\right] + 2 + \tilde\mu^2$ and $n'_3 = \left[\tilde\mu^3\right] + 1$ but for this weight, all states are null-states. Therefore, $N_{n'_i\alpha^{(i)}}=0$. In the end, we get $N_{n_i\alpha^{(i)}}=0$, for the above values of $n_i$. Therefore, unitarity excludes all representations with $\tilde\mu^2 > 0$, $\tilde{\mu}^3>-1$ and $\tilde{\mu}^3\notin\Z_+$ in all cases except $\mf{g}^{\C}=B_3$. In the case of $\mf{g}^{\C}=B_3$ this excludes $\tilde\mu^2 \geq 0$, $\tilde{\mu}^3>-1$ and $\tilde{\mu}^3\notin\Z_+$.

Let us summarize the results of this section in a theorem
\begin{theorem}
Let $\hat{\tilde{\mu}}$ satisfy $\tilde{\mu}^{l}>-1$ where $l=2,\ldots,r_{\mf{g}}$ and
\eqnb
\begin{array}{ll}
\mf{su}(r_{\mf{g}},1): & \tilde{k}  > \left(\tilde{\mu},{\theta'_{\mf{h}'}}^\vee\right) \\
\mf{so}(2r_{\mf{g}}-1,2): & \tilde{k} - 1 - \tilde{\mu}^2  > \left(\tilde{\mu},{\theta'_{\mf{h}'}}^\vee\right) \\
\mf{so}(2r_{\mf{g}},2): & \tilde{k} - 1 - \tilde{\mu}^2 > \left(\tilde{\mu},{\theta'_{\mf{h}'}}^\vee\right) \\
\end{array}
\nonumber
\eqne
and be one of the following cases:
\begin{enumerate}
\item[i)]{$\tilde{\mu}^2\notin\Z_+$.}
\item[ii)]{$\mf{g}^\C \neq B_3$, $\tilde\mu^2\in \Z_{+}\setminus{0}$ and $\tilde{\mu}^3\notin\Z_+$.}
\item[iii)]{$\mf{g}^\C = B_3$, $\tilde\mu^2\in \Z_{+}$ and $\tilde{\mu}^3\notin\Z_+$.}
\end{enumerate}
Then one can always find an antidominant $\hat{\mu}$ such that the state-space is non-unitary.
\end{theorem}

\subsection{Necessary conditions for $\mf{g}^\C=D_4$}
We consider here the case of $\mf{g}=\mf{so}(6,2)$ in more detail\footnote{As $\mf{so}^*(8)\cong \mf{so}(6,2)$, the conclusions hold also for this case.}. Let us limit our study to weights satisfying $\tilde{\mu}^2\in \Z_+$ and $\tilde{\mu}^l>-1$ for $l=3,4$. We first study some states in the auxiliary sector for non-unitary representations. The states which are of interest are of the form
\eqnb
\left|\phi \right> 
	&=&
		\left(\tilde{E}^{-\alpha^{(2)}}\right)^{\tilde{\mu}^2 + n_3 + n_4} \left(\tilde{E}^{-\alpha^{(3)}}\right)^{n_3} \left(\tilde{E}^{-\alpha^{(4)}}\right)^{n_4}\left|\tilde\mu,\tilde{k},0\right>
\eqne
which has degeneration one. The norm is
\eqnb
\left<\phi \left|\phi \right> \right.
	&=&
		\left(\left(\tilde{\mu}^2 + n_3 + n_4\right)!\right)^2\left(n_3\right)!\left(n_4\right)!
		\prod_{j_3=0}^{n_3 - 1} \left(\tilde{\mu}^3 - j_3\right) \prod_{j_4=0}^{n_4 - 1} \left(\tilde{\mu}^4 - j_4\right)
\eqne
which can be divided into three interesting cases
\eqnb
\left<\phi \left|\phi \right> \right.\hspace{-3mm}
	&=&
		\hspace{-3mm}
		\left\{
		\begin{array}{ll}
		 \hspace{-2mm} \left|K_1\right|^2\left(-\right)^{n_3 -\left[\tilde{\mu}^3\right] - 1} & \tilde{\mu}^{3,4} > -1\;, \tilde{\mu}^3\notin \Z,\; n_3 \geq \left[\tilde{\mu}^3\right] + 1, \; n_4 \leq \left[\tilde{\mu}^4\right] + 1 \\
		 \hspace{-2mm}\left|K_2\right|^2\left(-\right)^{n_4 -\left[\tilde{\mu}^4\right] - 1} & \tilde{\mu}^{3,4} > -1\;, \tilde{\mu}^4\notin \Z,\; n_4 \geq \left[\tilde{\mu}^4\right] + 1, \; n_3 \leq \left[\tilde{\mu}^3\right] + 1 \\
		 \hspace{-2mm}\left|K_3\right|^2\left(-\right)^{n_3 + n_4 - \left[\tilde{\mu}^3\right] - \left[\tilde{\mu}^4\right] - 2} & \tilde{\mu}^{3,4} > -1,\; \tilde{\mu}^{3,4}\notin \Z,\; n_3 \geq\left[\tilde{\mu}^3\right] + 1, \; n_4 \geq \left[\tilde{\mu}^4\right] + 1
		\end{array}
		\right. ,
		\no
\eqne
where cases one and two are interchanged by interchanging $\left(\tilde{\mu}^3,n_3\right)\leftrightarrow \left(\tilde{\mu}^4,n_4\right)$. $K_i$, $i=1,2,3$, are non-zero. From these equations we see that we have negative norms for  
\eqnb
\begin{array}{crclc}
1: & n_3 &=& \left[\tilde{\mu}^3\right] + 2l + 2, & l\in\Z_+ \\
2: & n_3 &=& \left[\tilde{\mu}^4\right] + 2l + 2, & l\in\Z_+ \\
3: & n_3 + n_4 &=& \left[\tilde{\mu}^3\right] + \left[\tilde{\mu}^4\right] + 2l + 3, & l\in\Z_+ \\
\end{array}
\eqne

We now study the term in eq.\ (\ref{Cond2.horizunitarity}) arising from the identity element of the Weyl group. It will be non-zero if
\eqnb
\tilde\mu^2 - \epsilon^2 + 1 + m &=& 2n_2 - n_3 - n_4 \no
\tilde\mu^3 - \epsilon^3 + 1     &=& - n_2 + 2n_3 \no
\tilde\mu^4 - \epsilon^4 + 1     &=& - n_2 + 2n_4
\label{so8eqn}
\eqne
Consider first the case where $\tilde{\mu}^4>0$ (or $\tilde{\mu}^3>0$, by symmetry). We can here set
\eqnb
m &=& \tilde{\mu}^2 + \left[\tilde{\mu}^3\right]+3 \no
\epsilon^2 &=& 1 \no
\epsilon^3 &=& \tilde{\mu}^2 + \tilde{\mu}^3 - \left[\tilde{\mu}^3\right] \no
\epsilon^4 &=& \tilde{\mu}^2 + \left[\tilde{\mu}^3\right] + \tilde{\mu}^4,
\eqne
corresponding to a highest weight for which the horizontal part is antidominant. The weight $\hat{\mu}$ is antidominant since the condition
\eqnb
\tilde{k}+2+\mu^1-\tilde{\mu}^2 &<& \left(\tilde{\mu},{\theta'}^\vee_{\mf{h}'}\right),
\eqne 
is satisfied since it is a weaker condition then case $iv)$ in Theorem \ref{theorem4}. The ansatz yields the solution
\eqnb
n_2 &=& \tilde{\mu}^2 + \left[\tilde{\mu}^3\right] + 3 \no
n_3 &=& \left[\tilde{\mu}^3\right] + 2 \no
n_4 &=& 1.
\label{eq617}
\eqne
The other Weyl transformations that can contribute are $w_3$, $w_4$ and $w_3w_4$. They give the same solution as above except that one shifts $(n_3,n_4)$ to $(\left[\tilde{\mu}^3\right] + 1,1)$, $(\left[\tilde{\mu}^3\right] + 2,0)$ and $(\left[\tilde{\mu}^3\right] + 1,0)$, respectively. Considering the representation theory of the auxiliary sector, all those reflections correspond to null-states. Therefore, $N_{n_i\alpha^{(i)}}=0$ in those cases. Then the integral given in eq.\ (\ref{Cond1.horizunitarity}) implies $N_{n_i\alpha^{(i)}}=0$ for the values of $n_i$ given in eq.\ (\ref{eq617}). This is a contradiction as $N_{n_i\alpha^{(i)}}\neq0$. Hence, we conclude that the state-space is non-unitary.

Let us now consider the case when $\tilde{\mu}^3\leq 0$ and $\tilde{\mu}^{4}\leq 0$. We set 
\eqnb
m &=& \tilde{\mu}^2 + \left[\tilde{\mu}^3\right] + 2 \no
\epsilon^{2} &=& 1 \no
\epsilon^{3} &=& \tilde{\mu}^2 + \tilde{\mu}^3-\left[\tilde{\mu}^3\right] - 1 \no
\epsilon^{4} &=& \tilde{\mu}^2 + \left[\tilde{\mu}^3\right] + \tilde{\mu}^4 + 3,
\eqne
which gives a weight $\hat{\mu}$ for which the horizontal part is antidominant if $\tilde{\mu}^2\geq1$. The weight $\hat{\mu}$ is antidominant if
\eqnb
\tilde{k}+\mu^1-\tilde{\mu}^2 &<& \left(\tilde{\mu},\theta'_{\mf{h}'}\right),
\eqne 
is satisfied. The ansatz yields the solution
\eqnb
n_2 &=& \tilde{\mu}^{2} + \left[\tilde{\mu}^3\right] + 2 \no
n_3 &=& \left[\tilde{\mu}^3\right] + 2  \no
n_4 &=& 0.
\eqne
The other Weyl transformations will not contribute as the corresponding states are null-states. This may be shown in the same way as for the previously considered weights. Therefore, the solution inserted into eq.\ (\ref{Cond1.horizunitarity}) yields a state-space which is non-unitary.

Let us summarize our results in a theorem.
\begin{theorem}
Consider $\mf{g}=\mf{so}(6,2)$ and weights $\hat{\tilde{\mu}}$ satisfying the conditions $\tilde{\mu}^{l}>-1$ for $l=2,3,4$ and $\tilde{k}-1-\tilde{\mu}^2 > \left(\tilde{\mu},\theta'_{\mf{h}'}\right)$. If the highest weight satisfies one of the following conditions
\begin{enumerate}
\item[i)]{$\tilde{\mu}^2\notin \Z_{+}$}
\item[ii)]{$\tilde{\mu}^2\in \Z_{+}$, $\tilde{\mu}^3>0$, $\tilde{\mu}^4>-1$ and $\tilde{\mu}^4\notin\Z_+$}
\item[iii)]{$\tilde{\mu}^2\in \Z_{+}$, $\tilde{\mu}^3>-1$, $\tilde{\mu}^3\notin\Z_+$ and $\tilde{\mu}^4>0$}
\item[iv)]{$\tilde{\mu}^2\in \Z_{+}\setminus\{0\}$; $\tilde{\mu}^4\notin\Z_+$ or $\tilde{\mu}^3\notin\Z_+$}
\end{enumerate}
Then there exists at least one antidominant weight $\hat{\mu}$ such that the state-space is non-unitary.
\end{theorem}


\sect{Discussion}
Our investigation of necessary conditions for unitarity for the gauged WZNW models considered here has been quite technical and involved. We have shown that at least for the general class of models where  $\mf{g}^\C = A_{r}$ it is possible to establish general necessary conditions for unitarity of the state-space which hold to any grade. These conditions are the same as the ones formulated in the previous paper restricted to non-zero weights. We have also studied further conditions arising from states at grade zero especially relevant for low rank algebras. Here we found more solutions. In particular, they allow for non-integer weights. However, we have not established that these solutions will survive at non-zero grade, except in one case, $\mf{h}'^\C = A_2$ and $\mf{g}^\C = A_{3}$ where these extra solutions are then excluded. It would, of  course, be interesting to study the equations at least at lower non-zero grades to see whether these new solutions survive for more general cases. This is straightforward, in principle, but the practical problem to analyze the equations may be quite difficult. The method used in section three to analyze the equations, can be applied to any one of the different algebras to determine necessary conditions for unitarity. However, the increased complexity for other algebras than the ones treated makes the problem quite difficult.


\vspace{0.5cm}

\noindent
{\bf{Acknowledgements}}
J.B.\ would like to thank the theory group at Karlstad University for the hospitality during the completion of this work. The work by J.B.\ is supported by the Swedish Research Council under project no.\ 623-2008-7048. S.H.\ is partially supported by the Swedish Research Council under project no.\ 621-2008-4129 and no.\ 621-2005-3424.

\appendix

\sect{Proof of Proposition \ref{characterexpansion}}
The proof is based on using the results eqs.\ (\ref{eq215}) and (\ref{eq216}) established in the proof of Theorem \ref{theorem1}. One needs to consider each of the three cases separately.

\paragraph{$\mf{g}=\mf{su}(p,1)$ and $\mf{h}'=\mf{su}(p)$:} We choose the following basis of the state-space
\eqnb
\left(E^{-\alpha_p}\right)^{n_p}\cdot\ldots\cdot\left(E^{-\alpha_1}\right)^{n_1}\left|\sigma\right>,
\eqne
where
\eqnb
\alpha_i &=& \sum_{j=1}^i\alpha^{(j)}.
\eqne
This is a basis as states of the form 
\eqnb
\prod_{i=2}^{p+1}\left(E^{-\alpha^{(i)}}\right)^{n_i}\left|\sigma\right>,
\eqne
are all null-states for the choice of $\sigma$.
To find the highest weights w.r.t.\ $\mf{h}'$ we project out states which do not satisfy $E^{\alpha^{(i)}}\left|\varphi\right> = 0$ for $i = 2,\ldots,p$. Starting by $i=p$ we find that $n_p = 0$. Assume that $n_i = 0$ for all $i \geq j$. Then it is easily shown that $n_{j-1}=0$ if $j\geq3$. Thus, by induction, it follows that the only highest weight states of $\mf{h}'$ in the representation are of the form
\eqnb
\left(E^{-\alpha_1}\right)^{n_1}\left|\sigma\right>,
\eqne
which have the weight $\mu = n_1\Lambda_{(2)} - \left(\sigma^1 + 2n_1\right)\Lambda_{(1)}$. Then
\eqnb
c_{\mu,n}&=& \delta_{n\Lambda_{(2)},\mu}.
\eqne
Inserting this into eq.\ (\ref{branchingfuncdecomp}) and using eq.\ (\ref{charg.simple}) gives
\eqnb
\frac{1}{\prod_{\alpha\in\Delta_n^+}\left(1-e^{-\ii\left(\theta,\alpha\right)-\ii\phi}\right)} 
	&=&
		\sum_{n=0}^{\infty}e^{-\ii n\phi}\frac{\sum_{w\in W_{\mf{h}'}}\sign(w)e^{\ii \left(\theta,w(n\Lambda_{(2)}+\rho_{\mf{h}'})-\rho_{\mf{h}'}\right)}}{\prod_{\alpha\in\Delta_c^+}\left(1-e^{-\ii\left(\theta,\alpha\right)}\right)}.
\eqne
This proves the first of the three cases.

\paragraph{$\mf{g}=\mf{so}(2p,2)$ and $\mf{h}'=\mf{so}(2p)$:} This case is slightly more complicated. Here a general state can be written in terms of a basis
\eqnb
\left(E^{-\alpha_{2p-2}}\right)^{n_{2p-2}}\cdot\ldots\cdot\left(E^{-\alpha_1}\right)^{n_1}\left|\sigma\right>
\label{eqnA12}
\eqne
where 
\eqnb
\alpha_i       &=& \sum_{j=1}^{i} \alpha^{(j)}\;\;\;i\leq p-1 \no
\alpha_p       &=& \sum_{j=1}^{p-2} \alpha^{(j)} + \alpha^{(p)} \no
\alpha_{p+1}   &=& \sum_{j=1}^{p} \alpha^{(j)} \no
\alpha_{p+1 + i} &=& \sum_{j=1}^{p} \alpha^{(j)} + \sum_{j=1}^{i} \alpha^{(p-1-j)}, \;\;\;i\leq p-3.
\label{eqa12}
\eqne
Consider now the requirement that this state has to have positive weight w.r.t.\ $H^i$ for $i=2,\ldots,p$. This requirement follows from eq.\ (\ref{branchingfuncdecomp}). Then
\eqnb
n_{i-1}-n_{i}-n_{2p-i}+n_{2p-1-i} &\geq& 0 \; ; \; 2 \leq i \leq p-2 \no
n_{p-2}-n_{p+1} - \left|n_p - n_{p-1}\right| &\geq& 0.
\eqne
Thus, we get the following conditions
\eqnb
n_1-n_{2p-2} \geq \ldots \geq n_{i-1}-n_{2p-i} \geq \ldots \geq n_{p-2}-n_{p+1} \geq \left|n_p - n_{p-1}\right| \geq 0.
\label{eqa14}
\eqne
This will restrict the number of simple roots in that can appear in eq.\ (\ref{eqnA12}). The number of simple roots are
\eqnb
N_1 &=& \sum_{j=1}^{2p-2} n_j\no
N_i &=& \sum_{j=i}^{2p-1-i}n_j + 2\sum_{j=1}^{i-1}n_{2p-1-j} \no
N_{p-1} &=& n_{p-1} + \sum_{j=1}^{p-2}n_{2p-1-j} \no
N_{p} &=& n_{p} + \sum_{j=1}^{p-2}n_{2p-1-j}.
\label{so(n,1)condition}
\eqne
Using eq.\ (\ref{so(n,1)condition}) it is straighforward to prove that
\eqnb
n_i &=& N_i - N_{i+1} + n_{2p-1-i}\ Ê\  i=2,\ldots , p-2.
\eqne
which by eq.\  (\ref{eqa14}) implies
\eqnb
N_{i} &\geq& N_{i+1}\phantom{N + N_{p}} 2\leq i\leq p-3 \no
N_{p-2} &\geq& N_{p-1} + N_{p}.
\eqne
We now study the existence of highest weight states w.r.t. $\mf{h}'$ in the representation. If $N_p$ (or $N_{p-1}$) is zero we directly see that the only possibility is $\left(E^{\alpha^{(1)}}\right)^n\left|\sigma\right>$, as it coincides with the $ \mf{su}(p,1)$ representation. Consequently, we need $N_p\neq 0$ and $N_{p-1}\neq0$ to get something non-trivial. We can always take $N_p\leq N_{p-1}$ due to the $\Z_2$ invariance of the Dynkin diagram. Then a general state may be written as
\eqnb
\left|\phi\right> 
	&=& \sum_{i_1=0}^{N_p}\sum_{i_2=0}^{i_1}\ldots\sum_{i_j=0}^{i_{j-1}}\ldots\sum_{i_{p-2}=0}^{i_{p-3}}C_{i_1,\ldots,i_{p-2}} \left(E^{-\alpha_{2p-2}}\right)^{N_{p}-i_1}\left(E^{-\alpha_{2p-3}}\right)^{i_1-i_2}\cdot\ldots
	\no
	&\times&
	\left(E^{-\alpha_{p+1}}\right)^{i_{p-3}-i_{p-2}}\left(E^{-\alpha_p}\right)^{i_{p-2}}\left(E^{-\alpha_{p-1}}\right)^{N_{p-1}-N_p+i_{p-2}}
	\no
	&\times&
	\left(E^{-\alpha_{p-2}}\right)^{N_{p-2}-N_{p-1}-N_{p}+i_{p-3}-i_{p-2}}\left(E^{-\alpha_{p-3}}\right)^{N_{p-3}-N_{p-2}+i_{p-4}-i_{p-3}}\cdot\ldots
	\no
	&\times&
	\left(E^{-\alpha_{2}}\right)^{N_{2}-N_{3}+i_{1}-i_{2}}\left(E^{-\alpha_{p-3}}\right)^{N_{1}+N_{p}-N_2-i_{1}}\left|\sigma\right>.
\eqne
Applying the highest weight condition w.r.t.\ $E^{\alpha^{(p)}}$ implies that  $C_{i_1,\ldots,i_{p-2}}$ and $C_{i_1,\ldots,i_{p-2}+1}$ are related. Therefore, the only independent coefficients are $C_{i_1,\ldots,i_{p-3},0}$. We now impose the highest weight condition w.r.t.\ $E^{\alpha^{(p-1)}}$. This gives $C_{i_1,\ldots,i_{p-3},0} = 0$ if $N_{p-1}>N_{p}$. Therefore, $N_p=N_{p-1}$, and in this case, $C_{i_1,\ldots,i_{p-3},0}$ and $C_{i_1,\ldots,i_{p-3}+1,0}$ are related, leaving only $C_{i_1,\ldots,i_{p-4},0,0}$ as independent. We continue by imposing the highest weight condition w.r.t. $E^{\alpha^{(p-2)}}$.  This implies $N_{p-2}=2N_{p}$ and that all coefficients depend on $C_{i_1,\ldots,i_{p-5},0,0,0}$. This can be continued to any simple root $\alpha^{(i)}$, $i=2,\ldots,p$, which in the end yields $N_i = 2N_p$ for $i=2,\ldots,p-2$. Thus, the only highest weight states w.r.t the subalgebra $\mf h'$ in the representation are of the form
\eqnb
\left(\left(E^{-\alpha_{2p-2}}\right)^{N_p}\left(E^{-\alpha_1}\right)^{N_1-N_p} + \ldots\right)\left|\sigma\right> \; \; N_1\geq 2N_p\geq0,
\eqne
where $\ldots$ denote correction terms needed to make a highest weight state. We can determine the branching function for these highest weight states
\eqnb
\Tr[e^{\left(\ii\phi H\right)}e^{\ii\left(\theta,H\right)}]
	&=& \sum_{n_1=0}^\infty\sum_{n_2=0}^{[n_1/2]}e^{-\ii\phi n_1}e^{\ii\theta_2\left(n_1-2n_2\right)} \no
	&=& \sum_{n_1=0}^{\infty}\sum_{n_2=0}^{\infty}e^{-\ii\phi\left(n_1+2n_2\right)}e^{\ii\theta_2n_1} \no
	&=& \frac{1}{1-e^{-2\ii\phi}}\sum_{n_1=0}^{\infty}e^{\left(\ii\theta^2-\ii\phi\right)n_1},
\eqne
where the factor involving the highest weight has been excluded. We can now finally insert this expression into eqs.\ (\ref{charg.simple}) and (\ref{branchingfuncdecomp})
\eqnb
\frac{1}{\prod_{\alpha\in\Delta_n^+}\left(1-e^{-\ii\left(\theta,\alpha\right)- \ii\phi}\right)} 
	&=&
		\frac{1}{1-e^{-2\ii\phi}}\sum_{n=0}^{\infty}e^{- \ii n\phi}\frac{\sum_{w\in W_{\mf{h}'}}\sign(w)e^{ \ii \left(\theta,w(n\Lambda_{(2)}+\rho_{\mf{h}'})-\rho_{\mf{h}'}\right)}}{\prod_{\alpha\in\Delta_c^+}\left(1-e^{-\ii\left(\theta,\alpha\right)}\right)}.
		\no
\eqne
Case three is, therefore, proven.

\paragraph{$\mf{g}=\mf{so}(2p+1,2)$ and $\mf{h}'=\mf{so}(2p+1)$:}
A basis of the state-space for this case is given by
\eqnb
\left(E^{-\alpha_{2p-1}}\right)^{n_{2p-1}}\cdot\ldots\cdot\left(E^{-\alpha_{1}}\right)^{n_{1}}\left|\sigma\right>,
\eqne
where
\eqnb
\alpha_i &=& \sum_{j=1}^i\alpha^{(j)},\;\; 0\leq i\leq p \no
\alpha_{p+i} &=& \sum_{j=1}^p\alpha^{(j)} + \sum_{j=1}^i\alpha^{(p+1-i)}, \;\; 1\leq i\leq p-1.
\eqne
Consider now the conditions which follow from that this state has to have positive weight w.r.t. $H^i$ for $i=2,\ldots,p$. This implies
\eqnb
n_{j-1}-n_{2p+1-j} &\geq& n_j - n_{2p-j}, \;\;\;\; 2\leq i\leq p-1 \no
n_{p-1}-n_{p+1}&\geq& 0,
\eqne
from which we get the conditions
\eqnb
n_1-n_{2p-1} \geq \ldots \geq n_j - n_{2p-j} \geq \ldots \geq 0.
\eqne
We will again study how this restricts the number of simple roots in the above basis. The number of simple roots are
\eqnb
N_i &=& \sum_{j=i}^{2p-i}n_j + 2\sum_{j=2p+1-i}^{2p-1}n_j.
\label{expansionso(2n-1,2)}
\eqne
One may again prove
\eqnb
n_i &=& N_i + N_{i+1} + n_{2p-i}.
\eqne
which implies the conditions
\eqnb
N_2\geq N_3 \geq \ldots \geq N_i \geq N_{i+1} \geq \ldots \geq N_p \geq 0.
\eqne
Next, we study the existence of highest weight states w.r.t. $\mf h'$ in this module. We need to separately consider the two different cases when $N_p$ is even and odd. Take $N_{p}$ odd and write $N_{p} = 2\tilde{n}_{p}+1$. Then a general state with a fixed weight can be written as
\eqnb
\left|\phi\right> 
	&=& \sum_{i_1=0}^{\tilde{n}_p}\sum_{i_2=0}^{i_1}\ldots\sum_{i_j=0}^{i_{j-1}}\ldots\sum_{i_{p-1}=0}^{i_{p-2}}C_{i_1,\ldots,i_{p-1}}
	\left(E^{-\alpha_{2p-1}}\right)^{\tilde{n}_p-i_1}\left(E^{-\alpha_{2p-2}}\right)^{i_1-i_2}\cdot\ldots
	\no
	&\times&
	\left(E^{-\alpha_{p+1}}\right)^{i_{p-2}-i_{p-1}}\left(E^{-\alpha_{p}}\right)^{1+2i_{p-1}}\left(E^{-\alpha_{p-1}}\right)^{N_{p-1}-2\tilde{n}_p-1+i_{p-2}-i_{p-1}}
	\no
	&\times&
	\left(E^{-\alpha_{p-2}}\right)^{N_{p-2}-N_{p-1}+i_{p-3}-i_{p-2}}\cdot\ldots\cdot\left(E^{-\alpha_{2}}\right)^{N_{2}-N_{3}+i_{1}-i_{2}}\left(E^{-\alpha_{1}}\right)^{N_{1}+\tilde{n}_p-N_{2}-i_{1}}\left|\sigma\right>.\no
	\label{stateexpansionso(2p+1)}
\eqne
By requiring the state to be highest weight w.r.t. $E^{\alpha^{(p)}}$ gives $C_{i_1,\ldots,i_{p-1}}=0$ for any $i_1,\ldots,i_{p-1}$. Consequently there do not exist any highest weight states w.r.t. $\mf h'$ for this case. For $N_{p}$ even, one may make an expansion as in eq.\ (\ref{stateexpansionso(2p+1)}). Here there exists highest weight states w.r.t.\ the compact subalgebra if $N_2=N_3=\ldots=N_{p-1}=N_p=2\tilde{n}_p$. Thus a highest weight state can be written as 
\eqnb
\left(\left(E^{-\alpha_{2p-1}}\right)^{\tilde{n}_p}\left(E^{-\alpha_{1}}\right)^{N_{1}-\tilde{n}_p}+\ldots\right)\left|\sigma\right>\;\;N_1 \geq 2\tilde{n}_p\geq 0
\eqne
where the dots are correction terms making the state a highest weight state w.r.t.\ the compact subalgebra. Thus, the branching function for these states is
\eqnb
\Tr[e^{\ii\left(\phi H\right)}e^{\ii\left(\theta,H\right)}]
	&=& \sum_{n_1=0}^{\infty}\sum_{n_2=0}^{[n_1/2]}e^{-\ii\phi n_1}e^{\ii\theta^2\left(n_1-2n_2\right)} \no
	&=& \sum_{n_1=0}^{\infty}\sum_{n_2=0}^{\infty}e^{-\ii\phi \left(n_1-2n_2\right)}e^{\ii\theta^2n_1} \no
	&=& \frac{1}{1-e^{-2\ii\phi}}\sum_{n_1=0}^{\infty}e^{\left(\ii\theta^2-\ii\phi\right)n_1}.
\eqne
Inserting this expression into eqs.\ (\ref{charg.simple}) and (\ref{branchingfuncdecomp}) proves case  two  in the proposition. This concludes the proof of proposition.$\Box$

\sect{Proof of Theorem \ref{charsigntheorem}}


In this appendix we will prove Theorem \ref{charsigntheorem}. Before focusing on this, we make some definitions and prove some lemmas which simplify the proof. Denote by $A_{j}$ the subalgebra of $\mf{h}'^\C \cong A_{r_\mf{g}-1}$ spanned by the elements $H^k$ for $k=2,\ldots,j+1$ and $E^{\pm\alpha}$ where $\alpha\in\Delta_+^c$ and is of the form $\alpha=\sum_{k=2}^{j+1} n_k\alpha^{(k)}$ for $n_k\in\Z_+$. The corresponding Weyl group of the algebra is denoted by $W_{j}$ and is generated by the elements $w_{(k)}$ for $2 \leq k \leq j+1$. Furthermore, we define $W_1=1$, where $1$ is the identity element. Let us also define the right product between elements of the Weyl group of $\mf{h}'^\C$ and $W_{j}$ by
\eqnb
W_{j}w = \left\{w'w: \forall w'\in W_{j}\right\}.
\eqne
Furthermore, define the weight $\gamma$ which satisfies
\eqnb
\left(\tilde{\mu}-\gamma,\Lambda_{(j)}\right)
&>&
\left\{
\begin{array}{ll}
\left(\tilde{\mu}^i+1\right)\delta_{i,j}              & i \leq j\leq r_{\mf{g}} \phantom{12} \tilde{\mu}^i\in\Z_+ \\
\left(\left[\tilde{\mu}^i\right]+2\right)\delta_{i,j} & i \leq j\leq r_{\mf{g}} \phantom{12} \tilde{\mu}^i\in\Z_+
\end{array}.
\right.
\eqne 
We can also choose an ordering of the operators such that all states can be represented by
\eqnb
\mathcal{U}\left(\mf{p}_{-}^1\right)\mathcal{U}\left(\mf{p}_{-}^2\right)\mathcal{U}\left(\mf{p}_{-}^3\right)\left|\tilde{\mu}\right>,
\eqne
modulo null-states. Here the set $\mathcal{U}\left(\mf{p}_{-}^1\right)$ is generated by the operators corresponding to negative roots in $A_{i-2}$.  $\mathcal{U}\left(\mf{p}_{-}^2\right)$ is 
\eqnb
\mathcal{U}\left(\mf{p}_{-}^2\right)
&=&
\left\{\prod_{j=1}^{i-1}\left(E^{-\alpha_j}\right)^{n_j}: \alpha_j=\sum_{k=1}^{j}\alpha^{(i+1-k)}\phantom{12} n_j\in\Z_+ \right\}.
\eqne
The set $\mathcal{U}\left(\mf{p}_{-}^3\right)$ is generated by operators  corresponding to negative roots of $\mf{h}'^{\C}$ not in $\mathcal{U}\left(\mf{p}_{-}^1\right)\cup\mathcal{U}\left(\mf{p}_{-}^2\right)$. Observe that if we have an element other then the unit element in $\mathcal{U}\left(\mf{p}_{-}^3\right)$ acting on the vacuum, the state has a weight less then $\gamma$. As we derive a character and signature function valid for states satisfying $\lambda>\gamma$ we can neglect such states and only consider the states which can be represented by
\eqnb
\mathcal{U}\left(\mf{p}_{-}^1\right)\mathcal{U}\left(\mf{p}_{-}^2\right)\left|\tilde{\mu}\right>.
\eqne

To simplify the proof of the theorem, we prove a few lemmas which is used in different steps in the proof. The first lemma is a recursion relation for the Weyl group. 

\begin{lemma}
For the Weyl group of $A_{p}$ we have the recursion relation 
\eqnb
W_{j-1} &\cong& W_{j-2}\cup \bigcup_{k=0}^{j-2}W_{j-2}\prod_{l=0}^{k}w_{(j-l)}
\eqne
\label{lemmaB1}
\end{lemma}

\paragraph{Proof:} The first step in the proof is to identify that the right-hand side is generated by $w^{(k)}$ for $2 \leq k \leq j$. Therefore, one only needs to prove that all elements of $W_{j-1}$ are generated. Denote $W^0=W_{j-2}$ and $W^{k+1}=W_{j-2}\prod_{l=0}^{k}w_{(j-l)}$ for $k=0,\ldots,j-2$. 
The second step in the proof is that we need to prove that $\dim\left( W^{k}\right)=(j-1)!$ for $0\leq k \leq j-1$. A trivial step is that $\dim\left(W^{0}\right)=(j-1)!$, as $W_{j-2}$ is the Weyl group of $A_{j-2}$. What we need to prove is that the mapping $\prod_{l=0}^{k}w_{(j-l)}:W_{j-2}\rightarrow W^{k+1}$ is a bijection. The mapping is by definition surjective and to prove that it is injective we need to prove that all elements in $W_{j-2}$ are mapped to different elements in $W^{k+1}$. We prove this by contradiction. Let $w,w'\in W_{j-2}$ where $w\neq w'$ which is assumed to be mapped to the same element in $W^{k+1}$
\eqnb
w \prod_{k=0}^{j-1}w_{(i-k)} &=& w' \prod_{k=0}^{j-1}w_{(i-k)}.
\eqne
We can multiply by the element $\prod_{l=j+1-k}^{j}w_{(l)}$, which is the inverse of $\prod_{l=0}^{k-1}w_{(j-l)}$, from the right to get $w=w'$, which is a contradiction. Thus, all elements in $W_{j-2}$ are mapped to different elements in $W^{k+1}$. Therefore, the mapping is a bijection. Thus, $\dim\left(W_{i-1}\right) = \dim\left(W^j\right) =(i-1)!$ for $0\leq j \leq i-1$.

The third step in the proof is to prove that $k = l$ iff $W^{k} = W^{l}$. The proof of the "if" case is trivial. Consider now the "only if" case, which is equivalent to ``If $k \neq l$ then $W^{k} \neq W^{l}$''. Let us first consider the case $k=0$. Let $w,w'\in W_{j-2}$, we should show that the equality 
\eqnb
w &=& w' \prod_{m=0}^{l-1}w_{(j-m)},
\eqne
is a contradiction. This can be verified by multiplying from the left by $w'$. Consider now $k\neq0$. We can assume that $k<l$ and consider the elements $1,w\in W_{j-2}$. Therefore, we should prove that the equality 
\eqnb
w \prod_{m=0}^{k-1}w_{(j-m)} &=& \prod_{m=0}^{l-1}w_{(j-m)}
\eqne
is a contradiction. Multiply from the right by the element $\prod_{m=j+1-k}^{j}w_{(m)}$ to get
\eqnb
w &=& \prod_{m=0}^{k-1}w_{(j-m)}w_{(j-k)}\prod_{m=j+1-k}^{j}w_{(m)}\prod_{m=k+1}^{l-1}w_{(j-m)}.
\label{eqB10}
\eqne
Let us use the properties of the Weyl group of $A_{p}$,
\eqnb
w_{k+1}w_{k}w_{k+1} &=& w_{k}w_{k+1}w_{k} \\
w_{k}w_{l} &=& w_{l}w_{k} \phantom{12345}k\neq l\pm 1,
\eqne
to rewrite eq.\ (\ref{eqB10}) as
\eqnb
w &=& \prod_{m=0}^{k-1}w_{(j-k+m)}\prod_{m=0}^{l-1}w_{(j-m)}
\eqne
denote by $w'=\prod_{m=0}^{k-1}w_{(j-k+m)}$, which is an element in $W_{j-2}$. We then have the same equation as in the case of $k=0$. Therefore, we have a contradiction.

Let us now prove the lemma. All elements on the right-hand side of the isomorphism are different, as shown above. The dimension of the set on the right-hand side is 
$\sum_{k=0}^{j-1}(j-1)! = j!$ which equals the dimension of $W_{j-1}$. As the left- and right-hand sides are furthermore generated by the same elements, they are isomorphic. $\Box$

\begin{lemma}
Let $w^{\{k\}}\in 1\cup\bigcup_{k=0}^{j-2}\prod_{l=0}^{k}w_{(j-l)}$ for $i \leq j\leq r_\mf{g}$ then
\eqnb
w^{\{k\}}\left(\tilde{\mu}+\rho_{\mf{h}'}\right)-\tilde{\mu}-\rho_{\mf{h}'}
&=&
\left\{
\begin{array}{ll}
0 & w^{(k)}=1 \\
-\left(\mu^{(i)} + 1\right)\alpha^{(i)} & w^{(k)} = w_{(i)} \\
\mathcal{O}\left(\gamma-\tilde{\mu}\right) & w^{(k)} \neq 1, \; w^{(k)} \neq w_{(i)}
\end{array}
\right.
\eqne
\label{lemmaB2}
\end{lemma}
\paragraph{Proof:} The case when $w^{\{k\}}=1$ is trivial as well as the case $w^{\{k\}} = w_{(i)}$. The latter is proven by
\eqnb
w_{(i)}\left(\tilde{\mu}+\rho_{\mf{h}'}\right)-\tilde{\mu}-\rho_{\mf{h}'}
    &=&
        -\left(\tilde{\mu}+\rho_{\mf{h}'},\alpha^{(i)^{\vee}}\right)\alpha^{(i)}
    \no
    &=&
        -\left(\tilde{\mu}^{i}+1\right)\alpha^{(i)}.
\eqne
Consider now the third case. Using the properties of the fundamental elements in the Weyl group for $A_p$,
\eqnb
w_{(i)}\left(\lambda\right) &=& \lambda + \lambda^{i}\alpha^{(i)} \no
w_{(j)}\left(\alpha^{(k)}\right) &=& -\alpha^{(j)} \phantom{123456} j=k\pm 1,
\eqne
we get
\eqnb
\lambda_{j,k} 
    &\equiv&
        \prod_{l=0}^{k}w_{(j-l)}\left(\tilde{\mu}+\rho_{\mf{h}'}\right)-\tilde{\mu}-\rho_{\mf{h}'}
    \no
        &=& -\sum_{m=j-k}^{j}\left[\sum_{l=j-k}^{m}\left(\tilde{\mu}^{l}+1\right)\right]\alpha^{(m)}.
\eqne
We can determine
\eqnb
\left(\tilde{\mu}-\lambda_{j,k},\Lambda_{(m)}\right) &=& 
\left\{
\begin{array}{ll}
\sum_{l=j-k}^{m}\left(\mu^{l}+1\right) & j-k\leq m \leq k \\
0. & 
\end{array}
\right.
\eqne
We proceed by proving the lemma by considering different values of $j$ and $k$. Consider first $j=i$. If we let $k=i-1$ we will get
\eqnb
\left(\tilde{\mu}-\lambda_{i,i-1},\Lambda_{(i)}\right) 
    &=&
        \left(\tilde{\mu}^{i} + 1 + \tilde{\mu}^{i-1}+1\right)
    \no
    &\geq&
        \left(\tilde{\mu}^{i} + 2\right)
    \no
    &>&
        \left(\tilde{\mu}-\gamma\right),
\eqne
thus $\lambda_{i,i-1} = \mathcal{O}(\gamma-\tilde{\mu})$ this holds as well for all $k\leq i-1$. Continuing, we look at the case $j\geq i+1$. Setting $m=i+1$ yields
\eqnb
\left(\tilde{\mu}-\lambda_{j,k},\Lambda_{(i)}\right)
    &=&
        \sum_{l=j-k}^{i}\left(\mu^{l}+1\right)
    \no
    &>& 0.
\eqne
Thus $\lambda_{j,k} = \mathcal{O}(\gamma-\tilde{\mu})$ if $j\geq i+1$. $\Box$

Let us prove a lemma which simplifies the analysis of which states at a certain weight, $\lambda$, satisfying $\left(\mu-\lambda,\Lambda_{(i)}\right)=N$, are highest weights of the subalgebra $A_{i-2}$. We should here consider the elements in $\mathcal{U}\left(\mf{p}^2_{-}\right)$ in more detail.
\begin{lemma}
Assume $j>k$. Let $\alpha = \sum_{l=k+1}^{j}\alpha^{(i+1-l)}$. Furthermore, denote by
\eqnb
E^{1} &=& E^{-\alpha_k} \no
E^{2} &=& E^{-\alpha_j} \no
E^{3} &=& E^{-\alpha}.
\eqne
Then we have the relation
\eqnb
\left(E^{2}\right)^{n_2}\left(E^{1}\right)^{n_1}
&=& \frac{n_1!}{(n_1+n_2)!}\sum_{j=0}^{n_2}\left(-1\right)^{j}\binom{n_2}{j}\left(E^{3}\right)^{n_2-j}\left(E^{1}\right)^{n_1+n_2}\left(E^{3}\right)^{j}.
\eqne
\label{lemmaB3}
\end{lemma}
\paragraph{Proof:} Let us prove this lemma by induction. The case $n_2=0$ is trivial. But we also need to consider the case $n_2=1$
\eqnb
E^2\left(E^1\right)^{n_1}
&=&
\left[E^3,E^1\right]\left(E^1\right)^{n_1}
\no
&=&
E^3\left(E^1\right)^{n_1+1}-E^1E^3\left(E^1\right)^{n_1}
\no
&=&
E^3\left(E^1\right)^{n_1+1}-\left(E^1\right)^{n_1+1}E^3-n_1E^2\left(E^1\right)^{n_1}.
\eqne
Therefore,
\eqnb
E^2\left(E^1\right)^{n_1} 
&=&
\frac{1}{n_1+1}\left(E^3\left(E^1\right)^{n_1+1}-\left(E^1\right)^{n_1+1}E^3\right)
\no
&=&
\frac{n_1!}{(n_1+1)!}\sum_{j=0}^{1}\left(-1\right)^{j}\binom{1}{j}\left(E^{3}\right)^{1-j}\left(E^{1}\right)^{n_1+1}\left(E^{3}\right)^{j},
\label{eqB28}
\eqne
which proves the case $n_2=1$.

Assume now that it is true for $n_2$ and consider the case $n_2+1$,
\eqnb
\left(E^{2}\right)^{n_2+1}\left(E^{1}\right)^{n_1} 
&=&
\frac{n_1!}{(n_1+n_2)!}\sum_{j=0}^{n_2}\left(-1\right)^{j}\binom{n_2}{j}\left(E^{3}\right)^{n_2-j}E^{2}\left(E^{1}\right)^{n_1+n_2}\left(E^{3}\right)^{j}.
\no
\eqne
Let us use eq.\ (\ref{eqB28}),
\eqnb
\left(E^{2}\right)^{n_2+1}\left(E^{1}\right)^{n_1}
&=&
\frac{n_1!}{(n_1+n_2+1)!}\sum_{j=0}^{n_2}\left(-1\right)^{j}\binom{n_2}{j}\left(E^{3}\right)^{n_2-j}
\no
&\times&
\left[E^{3}\left(E^{1}\right)^{n_1+n_2+1}-\left(E^{1}\right)^{n_1+n_2+1}E^{3}\right]\left(E^{3}\right)^{j}.
\eqne
Using the properties of binominal coefficients, $\binom{n_2}{j}+\binom{n_2}{j-1}=\binom{n_2+1}{j}$, one can simplify the equation to
\eqnb
\left(E^{2}\right)^{n_2+1}\left(E^{1}\right)^{n_1} &=& \frac{n_1!}{(n_1+n_2+1)!}\sum_{j=0}^{n_2+1}\left(-1\right)^{j}\binom{n_2+1}{j}
\no
&\times&
\left(E^{3}\right)^{n_2+1-j}\left(E^{1}\right)^{n_1+n_2+1}\left(E^{3}\right)^{j},
\eqne
which proves the lemma. $\Box$

Let us now prove a central lemma in the proof of the theorem.
\begin{lemma}
If $\tilde{\mu}^j\in\Z_+$ for $2\leq j \leq i-1$ and $\tilde{\mu}^i\notin \Z$ then all highest weights of the algebra $A_{i-2}$ at weights $\lambda$ satisfying $\left(\tilde{\mu}-\lambda,\Lambda_{(i)}\right)=N$ and $\left(\tilde{\mu}-\lambda,\Lambda_{(j)}\right)=0$ for $i+1 \leq j \leq r_{\mf{g}}$ are of the form
\eqnb
\left|n_j\right> &\equiv& \prod_{j=1}^{i-1}\left(E^{\alpha_j}\right)^{n_j}\left|\tilde{\mu}\right> + \ldots,
\eqne
where $\ldots$ denote descendants of other highest weights such that $\left|n_j\right>$ is a highest weight for $A_{i-2}$ and $\sum_{j=1}^{i-1}n_j=N$. Furthermore, $\mu^{k}\geq n_{i+1-k}$ for $2 \leq k \leq i-1$ and highest weights of $A_{i-2}$ are dominant.
\label{lemmaB4}
\end{lemma}
\paragraph{Proof:} We know that the highest weights satisfying the assumptions of the lemma are of the form 
\eqnb
\mathcal{U}\left(\mf{p}_{-}^2\right)\left|\tilde{\mu}\right> + \ldots.
\eqne
Therefore, we need to determine which of the states in $\mathcal{U}\left(\mf{p}_{-}^2\right)\left|\tilde{\mu}\right>$ satisfying $\left(\tilde{\mu}-\lambda,\Lambda_{(i)}\right)=N$ are true highest weights of $A_{i-2}$. The states $\mathcal{U}\left(\mf{p}_{-}^2\right)\left|\tilde{\mu}\right>$ are of the form
\eqnb
\prod_{j=1}^{i-1}\left(E^{-\alpha_j}\right)^{n_j}\left|\tilde{\mu}\right>,
\label{eqB34}
\eqne
and satisfy $\sum_{j=1}^{i-1}n_j=N$. As all operators commute in the above expression we can move the operator $E^{-\alpha_{k+1}}$ and the operator $E^{-\alpha_k}$ to the right. We can now use Lemma \ref{lemmaB3} for $j=k+1$ to get
\eqnb
\prod_{j=1}^{i-1}\left(E^{-\alpha_j}\right)^{n_j}\left|\tilde{\mu}\right>&=&
\prod_{m=1,m\neq k,m\neq k+1}^{i-1}\left(E^{-\alpha_m}\right)^{n_l}\left(E^{2}\right)^{n_{k+1}}\left(E^{1}\right)^{n_k}\left|\tilde{\mu}\right>
\no
&=&
\prod_{m=1,m\neq k,m\neq k+1}^{i-1}\left(E^{-\alpha_m}\right)^{n_m}\frac{n_k!}{(n_k+n_{k+1})!}
\no
&\times&
\sum_{l=0}^{n_{k+1}}\left(-1\right)^{l}\binom{n_{k+1}}{l}\left(E^{3}\right)^{n_{k+1}-l}\left(E^{1}\right)^{n_k+n_{k+1}}\left(E^{3}\right)^{l}\left|\tilde{\mu}\right>,
\no
\eqne
where we have used the notation introduced in Lemma \ref{lemmaB3}. As $\left(E^{3}\right)^{l}\left|\tilde{\mu}\right>$ is a null-state if $l\geq \tilde{\mu}^{(i-k)}+1$, the equation can be rewritten as 
\eqnb
&&
\prod_{m=1,m\neq k,m\neq k+1}^{i-1}\left(E^{-\alpha_m}\right)^{n_m}\left(E^{3}\right)^{\left(n_{k+1}-\tilde{\mu}^{(i-k)}\right)\delta_{n_{k+1} \geq \tilde{\mu}^{(i-k)}}}
\no
&&
\frac{n_k!}{(n_k+n_{k+1})!}\sum_{l=0}^{\tilde{\mu}^{(i-k)}}\left(-1\right)^{l}\binom{n_{k+1}}{l}\left(E^{3}\right)^{n_{k+1}-l}\left(E^{1}\right)^{n_k+n_{k+1}}\left(E^{3}\right)^{l}\left|\tilde{\mu}\right>,
\eqne
modulo null-states. Here $\delta_{a \geq 0}=1$ and $\delta_{a\geq0}=0$ if $a\geq 0$ and $a<0$, respectively. Assume first that $n_{k+1}\geq \tilde{\mu}^{(i-k)}+1$. As $E^{3}$ commutes with $\prod_{m=1,m\neq k,m\neq k+1}^{i-1}\left(E^{-\alpha_m}\right)^{n_m}$ we can rewrite the equation as
\eqnb
\prod_{m=1}^{i-1}\left(E^{-\alpha_m}\right)^{n_l}\left|\tilde{\mu}\right>
&=&
\sum_{p=0}^{\tilde{\mu}^{(i-k)}}a_p\left(E^{3}\right)^{\left(n_{k+1}+p-\tilde{\mu}^{(i-k)}\right)}\prod_{m=1,m\neq k,m\neq k+1}^{i-1}\left(E^{-\alpha_m}\right)^{n_m}
\no
&\times&
\left(E^{2}\right)^{\tilde{\mu}^{(i-k)}-p}\left(E^{1}\right)^{n_k+n_{k+1}+p-\tilde{\mu}^{(i-k)}}\left|\tilde{\mu}\right>,
\eqne
where we have used Lemma \ref{lemmaB3}. $a_p$ are undetermined coefficients, but can be determined using Lemma \ref{lemmaB3} iteratively. Therefore, if $n_{k+1}\geq \tilde{\mu}^{(i-k)}+1$, we can rewrite the state as a sum of descendants of other highest weight states. This is valid for $1 \leq k \leq i-2$. Thus, for the states given in eq.\ (\ref{eqB34}) to be highest weight states, they have to satisfy
\eqnb
n_{i+1-k}\leq \tilde{\mu}^{k},\phantom{123}2 \leq k \leq i-1.
\eqne
Consider now the weights of these states. Apply $H^k$ where $2 \leq k \leq i-1$ on the state yields
\eqnb
H^k\prod_{j=1}^{i-1}\left(E^{-\alpha_j}\right)^{n_j}\left|\tilde{\mu}\right>
&=&
\left(\mu^{k} + n_{i-l} - n_{i+1-k}\right)\prod_{j=1}^{i-1}\left(E^{-\alpha_j}\right)^{n_j}\left|\tilde{\mu}\right>
\no
&\geq&
\left(n_{i-l} \right)\prod_{j=1}^{i-1}\left(E^{-\alpha_j}\right)^{n_j}\left|\tilde{\mu}\right>.
\eqne Thus, the weights are dominant. We have now proven the lemma. $\Box$

\begin{lemma}
The state 
\eqnb
\prod_{j=1}^{i-1}\left(E^{-\alpha_j}\right)^{n_j}\left|\tilde{\mu}\right>,
\eqne
satisfying $\mu^{k}\geq n_{i+1-k}$ for $2 \leq k \leq i-1$, has the norm
\eqnb
\prod_{j=1}^{i-1}n_j!\left[\prod_{k_j=0}^{n_j-1}\left(\sum_{l=i+1-j}^i\tilde{\mu}^{l}-\sum_{l=i+2-j}^in_{i+1-j}-k_j\right)\right]
\eqne
\label{lemmaB5}
\end{lemma}
\paragraph{Proof:} Rearrange the operators as
\eqnb
\left|n_i\right> &=&\prod_{j=1}^{i-1}\left(E^{-\alpha_{i-j}}\right)^{n_{i-j}}\left|\tilde{\mu}\right>,
\eqne
and compute
\eqnb
\left<n_i\right.\left|n_i\right>
&=&
\left<\tilde{\mu}\right|\prod_{j=1}^{i-1}\left(E^{\alpha_{j}}\right)^{n_{j}}\prod_{j=1}^{i-1}\left(E^{-\alpha_{i-j}}\right)^{n_{i-j}}\left|\tilde{\mu}\right>.
\eqne
Due to the choice of ordering, this equals
\eqnb
\left<n_i\right.\left|n_i\right>
&=&
\left<\tilde{\mu}\right|\prod_{j=1}^{i-2}\left(E^{\alpha_{j}}\right)^{n_{j}}\left(E^{\alpha_{i-1}}\right)^{n_{i-1}-1}\left[E^{\alpha_{i-1}},\left(E^{\alpha_{i-1}}\right)^{n_{i-1}}\right]\prod_{j=2}^{i-1}\left(E^{-\alpha_{i-j}}\right)^{n_{i-j}}\left|\tilde{\mu}\right>
\no
&=&
\left<\tilde{\mu}\right|\prod_{j=1}^{i-2}\left(E^{\alpha_{j}}\right)^{n_{j}}\left(E^{\alpha_{i-1}}\right)^{n_{i-1}-1}\sum_{k=0}^{n_{i-1}-1}\left(E^{\alpha_{i-1}}\right)^{n_{i-1}-1-k}\sum_{l=2}^{i}H^l\left(E^{\alpha_{i-1}}\right)^{k}
\no
&\times&
\prod_{j=2}^{i-1}\left(E^{-\alpha_{i-j}}\right)^{n_{i-j}}\left|\tilde{\mu}\right>
\no
&=&
\left<\tilde{\mu}\right|\prod_{j=1}^{i-2}\left(E^{\alpha_{j}}\right)^{n_{j}}\left(E^{\alpha_{i-1}}\right)^{n_{i-1}-1}\left(E^{-\alpha_{i-1}}\right)^{n_{i-1}-1}\sum_{k=0}^{n_{i-1}-1}\left(\sum_{l=2}^{i}H^l-2k\right)
\no
&\times&
\prod_{j=2}^{i-1}\left(E^{-\alpha_{i-j}}\right)^{n_{i-j}}\left|\tilde{\mu}\right>
\no
&=&
\left<\tilde{\mu}\right|\prod_{j=1}^{i-2}\left(E^{\alpha_{j}}\right)^{n_{j}}\left(E^{\alpha_{i-1}}\right)^{n_{i-1}-1}\left(E^{-\alpha_{i-1}}\right)^{n_{i-1}-1}\left[n_{i-1}\left(\sum_{l=2}^{i}H^l-n_{i-1}+1\right)\right]
\no
&\times&
\prod_{j=2}^{i-1}\left(E^{-\alpha_{i-j}}\right)^{n_{i-j}}\left|\tilde{\mu}\right>.
\eqne
Proceeding iteratively yields
\eqnb
\left<\tilde{\mu}\right|\prod_{j=1}^{i-2}\left(E^{\alpha_{j}}\right)^{n_{j}}\left[n_{i-1}!\prod_{k=0}^{n_{i-1}-1}\left(\sum_{l=2}^{i}H^{l}-k\right)\right]\prod_{j=2}^{i-1}\left(E^{-\alpha_{i-j}}\right)^{n_{i-j}}\left|\tilde{\mu}\right>.
\eqne 
This can be simplified, by using
\eqnb
\sum_{l=2}^{i}H^{l}\prod_{j=2}^{i-1}\left(E^{-\alpha_{i-j}}\right)^{n_{i-j}}\left|\tilde{\mu}\right>
&=&
\left(\sum_{l=2}^{i}\tilde{\mu}^{l} + \sum_{j=2}^{i-1}n_{i-j}\left(\alpha_{i-j},\alpha_{i-1}\right)\right)
\prod_{j=2}^{i-1}\left(E^{-\alpha_{i-j}}\right)^{n_{i-j}}\left|\tilde{\mu}\right>
\no
&=&
\left(\sum_{l=2}^{i}\tilde{\mu}^{l}-\sum_{l=3}^{i}n_{i+1-l}\right)
\prod_{j=2}^{i-1}\left(E^{-\alpha_{i-j}}\right)^{n_{i-j}}\left|\tilde{\mu}\right>,
\eqne
to
\eqnb
n_{i-1}!\prod_{k=0}^{n_{i-1}-1}\left(\sum_{l=2}^{i}\tilde{\mu}^{l}-\sum_{l=3}^{i}n_{i+1-l}-k\right)
\prod_{j=2}^{i-1}\left<\tilde{\mu}\right|\prod_{j=1}^{i-2}\left(E^{\alpha_{j}}\right)^{n_{j}}\prod_{j=2}^{i-1}\left(E^{-\alpha_{i-j}}\right)^{n_{i-j}}\left|\tilde{\mu}\right>.
\eqne
This expression has the same form as the one previously studied. Therefore, one can proceed in the same way and, in the end, get the expression for the norm
\eqnb
\prod_{j=1}^{i-1}n_j!\left[\prod_{k_j=0}^{n_j-1}\left(\sum_{l=i+1-j}^i\tilde{\mu}^{l}-\sum_{l=i+2-j}^in_{i+1-j}-k_j\right)\right],
\label{eqB47}
\eqne
which proves the lemma. $\Box$

\begin{lemma}
If $\tilde{\mu}^{i}\in\Z_+$ and $N=\sum_{j=1}^{i-1} n_j \leq \tilde{\mu}^i+1$ then the only null-state among
\eqnb
\prod_{j=1}^{i-1}\left(E^{-\alpha_j}\right)^{n_j}\left|\mu\right>
\eqne
satisfying $\mu^{k}\geq n_{i+1-k}$ occurs when $n_j=\left(\tilde{\mu}^i+1\right)\delta_{j,1}$.
\label{lemmaB7}
\end{lemma}
\begin{lemma}
If $\tilde{\mu}^{i}\notin\Z_+$ and $N=\sum_{j=1}^{i-1} n_j\leq \left[\tilde{\mu}^i\right]+2$ then the only negatively normed state among
\eqnb
\prod_{j=1}^{i-1}\left(E^{-\alpha_j}\right)^{n_j}\left|\mu\right>
\eqne
satisfying $\mu^{k}\geq n_{i+1-k}$ occurs when $n_j=\left(\left[\tilde{\mu}^i\right]+2\right)\delta_{j,1}$.
\label{lemmaB6}
\end{lemma}
\paragraph{Proof:} As the two lemmas are similar, we prove them simultaneously. From eq.\ (\ref{eqB47}) one will get
\eqnb
\sum_{l=i+1-j}^i\tilde{\mu}^{l}-\sum_{l=i+2-j}^in_{i+1-j}-k_j
&=&
\tilde{\mu}^i - n_1 + \sum_{l=i+1-j}^{i-1}\left(\tilde{\mu}^{l}-n_{i+1-j}\right) + n_j -k_j
\no
&\geq&
\tilde{\mu}^i - n_1 + n_j - k_j.
\eqne 
Therefore, the factor in the eq.\ (\ref{eqB47}) which takes the smallest value, is for $j=1$ which gives a factor
\eqnb
\tilde{\mu}^i - k_1 \phantom{123} 0 \leq k_1 \leq n_1-1.
\eqne
Consider the case when $\tilde{\mu}^i$ is an integer. If $n_i \leq \tilde{\mu}^i$ then all states have positive norm. If $n_i = \tilde{\mu}^i+1$ then we have a null-state. 

Consider the case when $\tilde{\mu}^i$ is not an integer. If $n_i \leq \left[\tilde{\mu}^i\right]+1$ then all states have positive norm. If $n_i = \left[\tilde{\mu}^i\right]+2$ then this is the unique negatively normed state. $\Box$

Let us now prove the theorem. Consider first the case when $\tilde{\mu}^{i}\in\Z_+$. From the definition of $\gamma$ one gets that all the states satisfying $\lambda>\gamma$ are unitary. Therefore, the character and signature function are equal for states satisfying $\lambda>\gamma$. Furthermore, the character is given by the Weyl character formula for these weights. What one needs to study are which elements in the Weyl group of $\mf{h}'^{\C}$ one has to sum over. From the recursion relation in Lemma \ref{lemmaB1} and Lemma \ref{lemmaB2} the elements one needs to include in the summation are 
\eqnb
W_{i-2}\cup W_{i-2}w_{(i)}.
\eqne
As the sign of $w_{(i)}$ is equal to minus one, and acts on $\tilde{\mu}+\rho_{\mf{h'}}$ as
\eqnb
w_{(i)}\left(\tilde{\mu} + \rho_{\mf{h'}}\right) 
&=&
\tilde{\mu} + \rho_{\mf{h'}} - \left(\left[\tilde{\mu}^{i}\right]+1\right)\alpha^{(i)}, 
\eqne 
we have proven the theorem for $\tilde{\mu}^{(i)}\in\Z$. 

Consider now $\tilde{\mu}^{i}\notin\Z_+$. Let us first study states which satisfy $\left(\tilde{\mu}-\lambda,\Lambda_{(i)}\right) \leq \left(\left[\tilde{\mu}^{i}\right]+1\right) < \left(\tilde{\mu}^{i}+1\right)$. From Lemma \ref{lemmaB4} we know that all states at this level can be written as descendents of dominant integer highest weights of the form
\eqnb
\prod_{k=1}^{i-1}\left(E^{-\alpha_k}\right)^{n_k}\left|\tilde{\mu}\right>.
\eqne
From Lemma \ref{lemmaB7} we know that all states are unitary. Lemma \ref{lemmaB1} and Lemma \ref{lemmaB2} imply that the elements one needs to include in the summation are equal to $W_{i-2}$. Thus, we have proven the theorem for $\tilde{\mu}^{i}\notin\Z_+$ for states satisfying $\left(\tilde{\mu}^{i}-\lambda,\Lambda_{(i)}\right) \leq \left(\left[\tilde{\mu}^{i}\right]+1\right)$. 

We proceed by studying states satisfying $\left(\tilde{\mu}-\lambda,\Lambda_{(i)}\right) = \left(\left[\tilde{\mu}^{i}\right]+2\right)$. Lemma \ref{lemmaB5} tells us that we do not have any new, fundamental, null-states at this level. The only difference is that one of the highest weight states embedded has negative norm, see Lemma \ref{lemmaB6}. Therefore, the character which is valid for weights $\left(\tilde{\mu}^{i}-\lambda,\Lambda_{(i)}\right) \leq \left(\left[\tilde{\mu}^{i}\right]+1\right)$ is also valid for $\left(\tilde{\mu}^{i}-\lambda,\Lambda_{(i)}\right) \leq \left(\left[\tilde{\mu}^{i}\right]+2\right)$ as well. For the signature function the only difference is that one of the highest weight states has negative norm. As this is a dominant integer highest weight, all descendants have negative norms. By using the Gram-Schmidt procedure and Lemma \ref{lemmaB6}, one finds that these states are the only non-unitary ones. Thus, to get the signature we need to subtract twice the contribution from these states 
\eqnb
-2\frac{\sum_{w\in W_{i-2}}\sign\left(w\right)e^{\ii\left(w\left(\tilde{\mu}+\rho_{\mf{h}'}-\sum_{j=i}^{r_{\mf{g}}}\left(\tilde{\mu}^{j}+1\right)\Lambda_{(j)}+\left(\left[\tilde{\mu}^{i}\right]+2\right)\Lambda_{(i-1)}\right)-\rho_{\mf{h}'}-\sum_{j=i}^{r_{\mf{g}}}\Lambda_{(i)},\theta\right)}}{\prod_{\alpha\in \Delta_+^{(i-2)}}\left(1-e^{-\ii\left(\alpha,\theta\right)}\right)}
\label{eqb51}
\eqne
where $\Delta_+^{(i-2)}$ is the positive roots for the algebra $A_{i-1}$. Using
\eqnb
w\left(\Lambda_{(j)}\right)
&=&
\Lambda_{(j)} \phantom{\mathcal{O}(e^{\ii\left(\gamma,\theta\right)})} i\leq j \leq r_{\mf{g}}\;\;w\in W_{i-2}
\\
w\left(\Lambda_{(i-1)}\right)
&=&
-w\left(\alpha^{(i)}\right) \phantom{\mathcal{O}(e^{\ii\left(\gamma,\theta\right)})} w\in W_{i-2}
\\
\frac{e^{\ii\left(\mu-\left(\left[\tilde{\mu}\right]+2\right)\alpha^{(i)},\theta\right)}}{\prod_{\alpha\in\Delta^c_+\setminus\Delta^{(i-1)}_+}\left(1-e^{-\ii\left(\alpha,\theta\right)}\right)}
&=&
\mathcal{O}(e^{\ii\left(\gamma,\theta\right)}),
\eqne
we can simplify eq.\ (\ref{eqb51}) to
\eqnb
-2e^{\ii\left(\tilde{\mu},\theta\right)}\frac{\sum_{w\in W_{i-2}}\sign\left(w\right)e^{\ii\left(w\left(\tilde{\mu}+\rho_{\mf{h}'}-\left(\left[\tilde{\mu}\right]+2\right)\alpha^{(i)}\right)-\tilde{\mu}-\rho_{\mf{h}'},\theta\right)}}{\prod_{\alpha\in \Delta_+^{c}}\left(1-e^{-\ii\left(\alpha,\theta\right)}\right)}.
\eqne
Therefore, if we add this term to the character one gets the signature function in the theorem. This proves the theorem. $\Box$

\end{document}